\documentclass{aa}  
\usepackage{comment}

\usepackage[utf8]{inputenc}
\usepackage[dvipsnames]{xcolor}
\usepackage{soul} 
 
\usepackage{txfonts}

\usepackage{graphicx}	
\usepackage{amsmath}	
\usepackage{amssymb}	
\usepackage{algorithm} 
\usepackage{algpseudocode} 
\usepackage{bm}
\usepackage[colorlinks=true,
    linkcolor=blue,
    citecolor=blue,
    filecolor=magenta,      
    urlcolor=cyan]{hyperref}

\newcommand{\lsim}{\mathrel{\rlap{\lower 3pt \hbox{$\sim$}} \raise 2.0pt \hbox{$<$}}}
\newcommand{\gsim}{\mathrel{\rlap{\lower 3pt \hbox{$\sim$}} \raise 2.0pt \hbox{$>$}}}

\newcommand{\fr}[1]{\textcolor{black}{ #1}}

\begin{document} 

   \title{BANG-MaNGA: A census of kinematic discs and bulges across mass and star formation in the local Universe} 
   \titlerunning{BANG-MaNGA}
   \authorrunning{F. Rigamonti et al.}

   \author{Fabio Rigamonti
          \inst{1,}
          \inst{2,} \inst{3}\fnmsep\thanks{frigamonti@uninsubria.it},
          Luca Cortese
          \inst{5,}
          \inst{6},
          Francesco Bollati
          \inst{1,}
          \inst{3},
          Stefano Covino
          \inst{2},
          Massimo Dotti
          \inst{2,}
          \inst{3,} \inst{4},
          A. Fraser-McKelvie
          \inst{6,}
          \inst{7,}
          \and
          Francesco Haardt
          \inst{1,}
          \inst{2,}
          \inst{3}
    }

   \institute{
            DiSAT, Universit\`a degli Studi dell'Insubria, via Valleggio 11, I-22100 Como, Italy
        \and 
            INAF - Osservatorio Astronomico di Brera, via Brera 20, I-20121 Milano, Italy
        \and
            INFN, Sezione di Milano-Bicocca, Piazza della Scienza 3, I-20126 Milano, Italy
        \and
            Universit\`a degli Studi di Milano-Bicocca, Piazza della Scienza 3, 20126 Milano, Italy
        \and
            International Centre for Radio Astronomy Research, The University of Western Australia, Crawley, WA 6009, Australia
        \and
            ARC Centre of Excellence for All Sky Astrophysics in 3 Dimensions (ASTRO 3D), Australia
        \and
            European Southern Observatory, Karl-Schwarzschild-Stra{\ss}e 2, 85748, Garching, Germany\\
    }

   \date{Received XXX; accepted YYY}

\abstract{
In this work, we aim to quantify the relevance of kinematically identified bulges and discs and their role in the process of galaxy quenching. To achieve this, we utilize an analysis of the SDSS-MaNGA survey conducted with the GPU-based code \textsc{bang} which simultaneously models galaxy photometry and kinematics to decompose galaxies into their structural components. We found that below $M_{\star}\simeq10^{11}~M_{\odot}$, galaxies span a wide range in their dynamical properties. The overall dynamical state of a galaxy is determined by the relative prominence of a dispersion-supported inner region and a rotationally-supported disc. Our decomposition reveals a natural separation between these classes, with only a minor fraction of stellar mass retained by structures exhibiting intermediate dynamical support. When examining galaxies in terms of their star formation activity, an apparent substantial decrease in rotational support is observed as they move below the star-forming main sequence. This behaviour is particularly evident when using luminosity-weighted tracers of kinematics, while it almost vanishes with mass-weighted tracers. Luminosity-weighted quantities not only capture differences in kinematics but also in the stellar population, potentially leading to biased interpretations of galaxy dynamical properties and quenching. Our findings indicate that quenching does not imply almost any structural transformation in galaxies below $M_{\star}\simeq10^{11}~M_{\odot}$. Processes such as disc fading are more likely explanations for the observed differences in mass-weighted and luminosity-weighted galaxy properties. Indeed, when the galactic disc ceases star formation, its mass-to-light ratio increases without any significant morphological transformation. The picture is remarkably different above $M_{\star}\simeq10^{11}~M_{\odot}$. In this case, regardless of the tracer used, a substantial increase in galaxy dispersion support is observed along with a significant structural change. A different quenching mechanism, most likely associated with mergers, dominates. Notably, this mechanism is confined to a very limited range of high masses.
}


\maketitle

\section{Introduction}
The morphological complexity of galaxies in the local Universe exemplified by the Hubble tuning fork \citep[][]{Hubble_1926}{}{}, and the strong correlation between morphology and other galaxy properties such as star formation, mass and kinematics  \citep[][]{Cappellari_2011,Kormendy_2012}{}{} , suggest that galaxies do grow following a variety of evolutionary pathways.

Multiple solutions have been proposed to shed light on the main formation mechanisms in shaping local galaxies; with a lot of attention given to techniques able to separate galaxies into different structural components identified from 2D imaging \citep[i.e. photometrically identified bulges/spheroid and discs,][]{Peng_2002,Gadotti_2009,Simard_2011,Lange_2016}{}{}. 
The preliminary studies about the relative importance of bulges and discs lead to the formulation of the "two phases" evolutionary framework \citep[][]{Oser_2010,Driver_2013}{}{}. In this commonly accepted scenario, bulges are the result of violent and rapid processes that happened relatively early in the Universe while discs are the result of accretion and cooling of gas conserving angular momentum \citep[][]{Fall_1980}. Moreover, galaxies in high-mass halos tend to build most of their mass "ex-situ" with little ongoing star formation at low redshift. In contrast, in the lower mass regimes, "in-situ" star formation is still present even at later stages. In principle, a "classical bulge" should be defined as the central galaxy component which has formed at early times as a result of violent processes. In practice, since it is impossible to know exactly the galaxy evolution history, this definition is never adopted. From here on, as we state again in Sec.~\ref{sec:data_sample}, we refer to dispersion-supported bulges as the galaxy light component which is spherically symmetric, non-rotating and isotropic\footnote{Note that these are exactly the same properties of the bulge component adopted in \textsc{bang}.}. 

Addressing the relative importance of bulges and discs in the local Universe is crucial for understanding the most relevant mechanism in forming galaxies. Results coming from different, purely photometric, decomposition approaches broadly agree equally splitting stellar mass into the bulge/spheroid and disc components \citep[][]{Moffet_2016,Thanjavur_2016,Bellstedt_2023}{}{}, with spheroidal galaxies being the dominant class above $M_{\star}\simeq 10^{11}\rm M_{\sun}$, contributing up to 80\%-100\% of the total mass budget. Even though these results may hint toward a Universe in which mergers play a prominent role \citep[][]{van_Dokkum2010,Lopez_2012}{}{} possible biases may exist. For instance, purely photometric decomposition approaches struggle in disentangling between nearly face-on dynamically cold discs and purely triaxial spheroid. Additionally, it has been shown that the commonly adopted definition of photometric bulges \citep[i.e. 'the central light component that is in excess of the inward extrapolation of an exponential fit to the disc brightness profile'][]{Kormendy_2004}{}{} may result in the inclusion of structures (i.e. nuclear disc, inner bars, pseudobulges) shaped by secular evolution processes \citep[][]{Erwin_2015,Erwin_2021,Gadotti2020}{}{}. In this regard, recent progress has been made through integral field spectroscopy (IFS) surveys, which have allowed us to classify galaxies according to their rotational-to-dispersion velocity ratio \citep[][]{Cappellari_2007}{}{}, showing that dispersion-dominated systems (or slow rotators) contribute minimally to the global mass budget in the local Universe \citep[][]{van_de_Sande_2017,Guo_2020}{}{}. 

In this regard, the galaxy decomposition problem has been approached also from an IFS perspective. This has been done either by slicing datacubes into several narrow bands and applying bulge+disc decomposition on them \citep[][]{Johnston_2017,Mendez2019}{}{} or by assuming a photometric decomposition and extracting the bulge and disc spectra directly from the datacubes \citep[][]{Oh_2016,Tabor_2017,Tabor_2019,Pak_2021}. These methods mostly focus on characterizing the stellar population and the star formation histories of bulges and discs without aiming at any dynamical models of the galaxy. An alternative approach is to model the galaxy surface brightness and its line-of-sight kinematics as the superposition of purposely weighted orbital families \citep[][]{Schwarzschild1979}. In this case, galaxies are characterized in terms of their angular momentum ex-post the fit without explicitly discriminating between bulges and discs. Without doubt, it is clear that linking visual morphology to the galactic stellar orbital distribution is not straightforward \citep[][]{Zhu_2018b,Fraser-McKelvie_2022}{}{}.

A proper characterization of galaxy structure becomes even more important when we are interested in linking morphology to the current star formation rate. It is generally assumed that the bimodality in SFRs \citep[][]{Brinchmann_2004}{}{} is closely linked to the structural and morphological properties of galaxies. This comes from the evidence that star formation mostly happens in disc-like galaxies, while spheroids and bulge-dominated galaxies are almost always quiescent. 

This oversimplified picture would suggest that the processes involved in quenching galaxies (i.e. shutting their star formation off) and changing their morphology (from a disc-like to a spheroidal shape) are closely related and may even act on similar timescales. However, the existence of a significant population of passive, rotationally dominated, disc galaxies has shown how this picture is too simplistic and that quenching may not necessarily imply morphological transformation. \citep[e.g.][]{Blanton2005,Cortese2009,Woo2017,Rizzo2018,Fraser-McKelvie_2021,Cortese_2022}{}{}. In the case of dissipationless (i.e. dry) major mergers, a significant alteration of the galaxy morphology toward a spheroidal shape is indeed expected \citep[][]{Cox2006}{}{}. Although, in the presence of gas a disc can be reformed after the merger. Similarly, it is argued that the growth of a central spheroidal structure can stabilize the gas disc preventing further fragmentation, star formation and quenching the galaxy \citep[][]{Martig2009}{}{}\footnote{This process is also known as "morphological quenching"}. However, such a mechanism implies the existence of passive, bulge-dominated galaxies surrounded by large gaseous discs a population which have not been, so far, observed. An apparent growth of a bulge related to the halting of star formation can be explained as a disc fading \citep[][]{Croom2021}{}{}. In this case, the galactic disc ceases star formation and dims, artificially boosting the galaxy bulge-to-total ratio. 

In \cite{Rigamonti_2022}, hereafter Paper I, we presented \textsc{bang}\footnote{\url{https://pypi.org/project/BANGal/}}, a code purposely designed for a simultaneous and self-consistent bulge+disc decomposition of galaxy photometry and kinematics. Being informed by the kinematics \textsc{bang}'s decomposition suffers minimally from the possible biases of purely photometric decomposition.
\textsc{bang} bases its parameter estimation on a robust Bayesian algorithm \citep[i.e. nested sampling][]{Skilling2004}{}{} combined with a GPU parallelization strategy for high computational performance, representing an optimal tool for automated analysis on large galaxy samples. This allowed us to present the first morpho-kinematic bulge+disc decomposition obtained through galaxy dynamical models \citep[][hereinafter Paper II]{Rigamonti_2023}{}{}
of the largest, to date, IFS survey \citep[i.e. the Mapping Nearby Galaxies at Apache Point Observatory, MaNGA, survey][]{Bundy_2015}{}{}.

In the present study, starting from the analysis of Paper II, we aim at characterising the mass budget of dispersion-supported bulges and discs with a particular emphasis on their kinematic state. Furthermore, by characterising the global kinematic state of each galaxy, we will measure the relative importance of dispersion-dominated and rotationally-dominated galaxies demonstrating that the inclusion of kinematics in the analysis together with proper dynamical modelling of galaxies is critical to reduce the biases present in purely photometry-based approaches. In addition, by including SFRs in our analysis, we will show how luminosity-weighted quantities may lead to improper conclusions on the importance of morphological transformation during the quenching phase of galaxies. Indeed, our results hint toward a morphological transformation almost not happening during quenching (at least below $M_{\star} \simeq10^{11}~M_{\sun}$) suggesting that processes such as bulge growth may not be the main cause of halting galaxies star formation. 

This paper is organised as follows. In \S2 we briefly set out the sample and the parameter catalogue we consider in this analysis together with a brief summary of our modelling approach. \S3 discusses in detail the role of some parameters that play a key role in the analysis of galaxies mass budget which is presented in \S4. \S5 and \S6 are devoted to a discussion and a summary of the interpretations and implications of our results. 

Throughout this paper, we use a flat $\Lambda$
cold dark matter cosmology: $H_0 = 70~\rm km ~ s^{-1} ~ Mpc^{-1}$ , $\Omega_M = 0.3$, and $\Omega_{\Lambda} = 0.7$.


\section{Data Sample and Parameters}
\label{sec:data_sample}

\begin{table}
	\centering
	\caption{Summary of the model parameters. The two columns describe each parameter and its reference name used in this work. When not otherwise specified, positions and lengths, masses, angles, and mass-to-light ratios are in physical units of $\rm kpc$, $\rm M_{\sun}$, degrees, and $\rm M_{\sun}/L_{\sun}$.}
	\label{tab:parameters}
	\small{\begin{tabular}{llcccc} 
        \hline\vspace{-0.75em}\\
		Description & Name\vspace{0.2em}\\
        \hline\vspace{-0.75em}\\
		Horizontal/vertical position of the centre & $x_0$ ; $y_0$ \\
		Position angle & $\mathrm{P.A.}$  \\
		Inclination angle & $i$  \\
		Bulge mass & $\log_{10}{M_{\rm b}}$ \\
  		Bulge scale radius & $\log_{10}{R_{\rm b}}$ \\
    	  Bulge inner slope & $\gamma$  \\
		Inner disc mass & $\log_{10}{M_{{\rm d},1}}$ \\
		Inner disc scale radius & $\log_{10}{R_{{\rm d},1}}$  \\
		Outer disc mass & $\log_{10}{M_{{\rm d},2}}$ \\
		Outer disc scale radius & $\log_{10}{R_{{\rm d},2}}$  \\
		Halo-to-stellar mass fraction & $log_{10}{f_{\star}}$  \\
		Halo concentration & $c$   \\
		mass-to-light ratio of the bulge & $M/L_{\rm b}$ \\
		Mass-to-light ratio of the inner disc & $M/L_{{\rm d},1}$ \\
		  Mass-to-light ratio of the outer disc & $M/L_{{\rm d},2}$ \\
		Kinematical decomposition parameter of the inner disc & $k_1$ \\
		Kinematical decomposition parameter of the outer disc & $k_2$ \vspace{0.2em}\\
		  \hline
	\end{tabular}}
\end{table}

\subsection{Photometry and Kinematics}
This study is based on a morpho-kinematic dataset built upon the SDSS imaging survey \citep[][]{SDSS} and the final release of the MaNGA survey from the Seventeenth Data Release \citep[DR17,][]{Abdurro'uf_2022} of the fourth phase of the Sloan Digital Sky Survey \citep[SDSS-IV,][]{Blanton_2017}. The MaNGA survey \citep{Bundy_2015,Drory_2015} has collected integral field spectroscopic measurements for $\simeq10,000$ local galaxies selected to have a roughly flat log stellar mass distribution above $10^{9}~\rm M_{\sun}$ in the redshift range $0.01 < \rm z <0.15$. In this work, we use line-of-sight (l.o.s) stellar velocity and velocity dispersion estimates and their associated errors as provided by the data analysis pipeline \citep[\textsc{dap},][]{Westfall_2019}{}{} assuming the same native spatial binning as for the data cube.
On top of that, we collected SDSS $i$ and $g$ band surface brightness maps and cropped them to the same spatial extent as the MaNGA data.

\begin{figure*}
    \centering
    \includegraphics[scale=0.37]{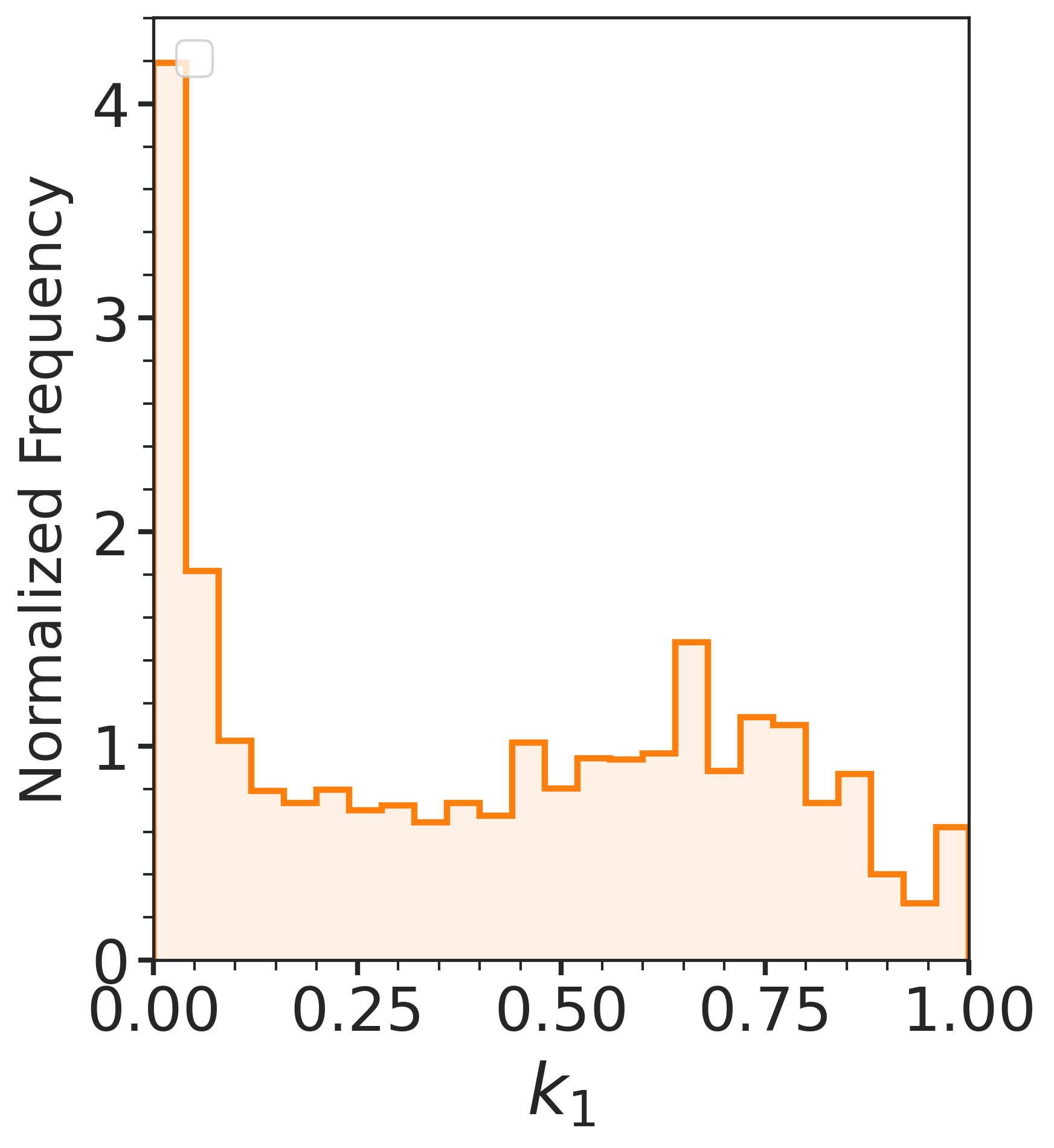}\hspace{0.01\textwidth}
    \includegraphics[scale=0.37]{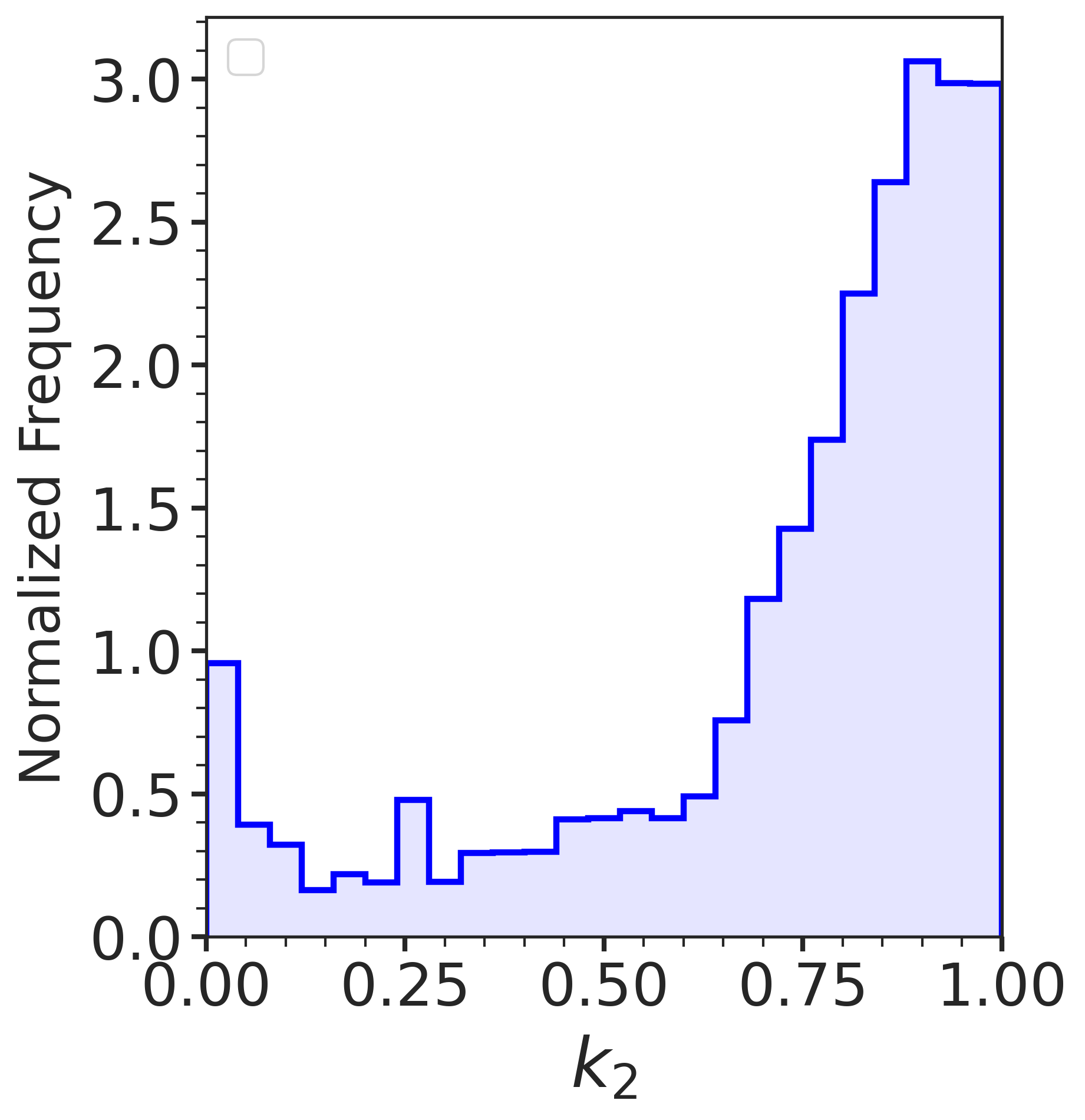}\hspace{0.01\textwidth}
    \includegraphics[scale=0.37]{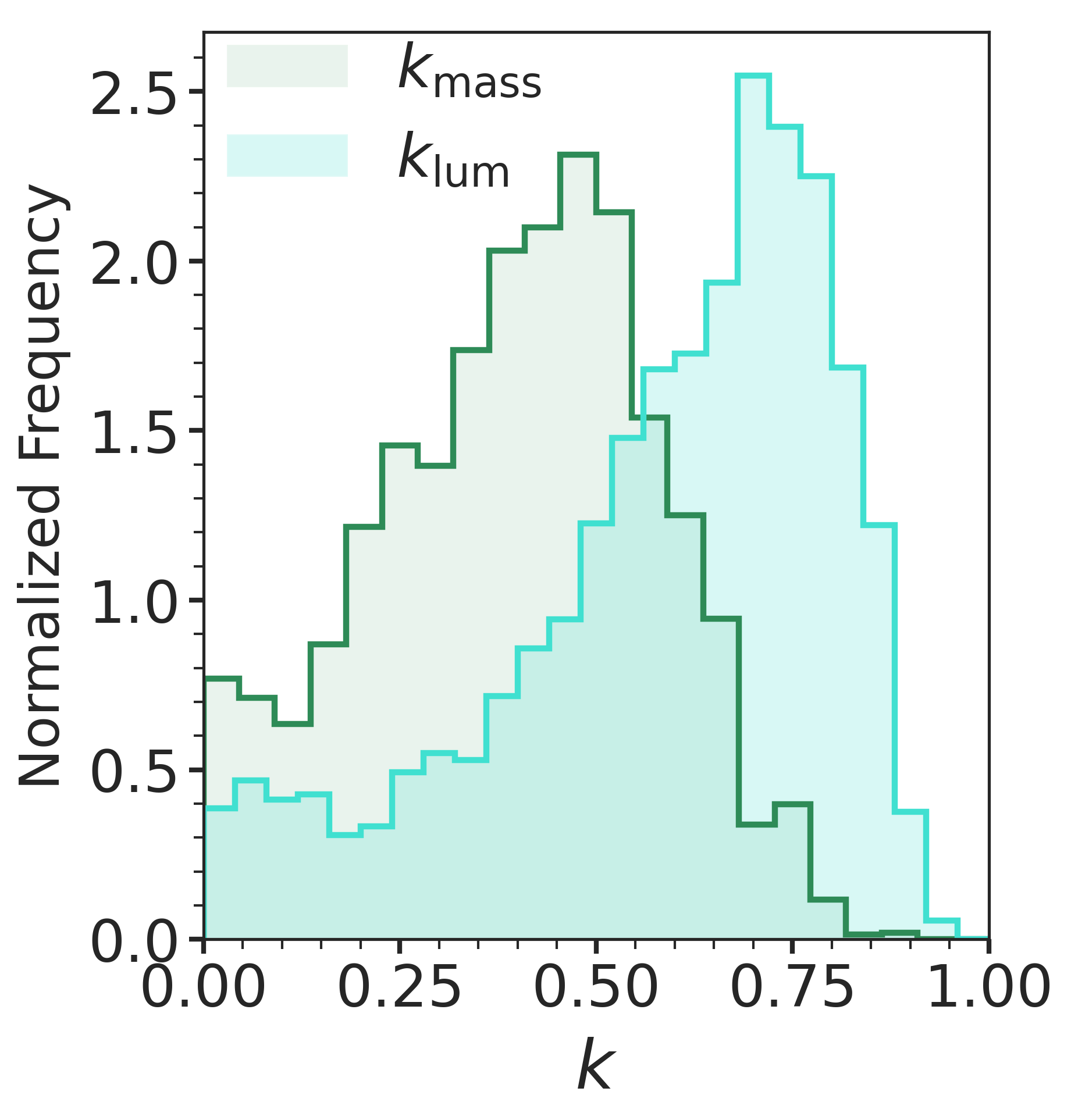}\hspace{0.01\textwidth}
    \caption{From left to right: normalized volume-weighted distributions of the $k_1$, $k_2$, $k_{\rm lum}$ (in turquoise) and $k_{\rm mass}$ (in dark green) parameters. See Eq.~\ref{eq:k_lum} and Eq.~\ref{eq:k_mass} for the definitions of $k_{\rm lum}$ and $k_{\rm mass}$.}
    \label{fig:k_distributions}
\end{figure*}

\begin{figure*}
    \includegraphics[scale=0.304]{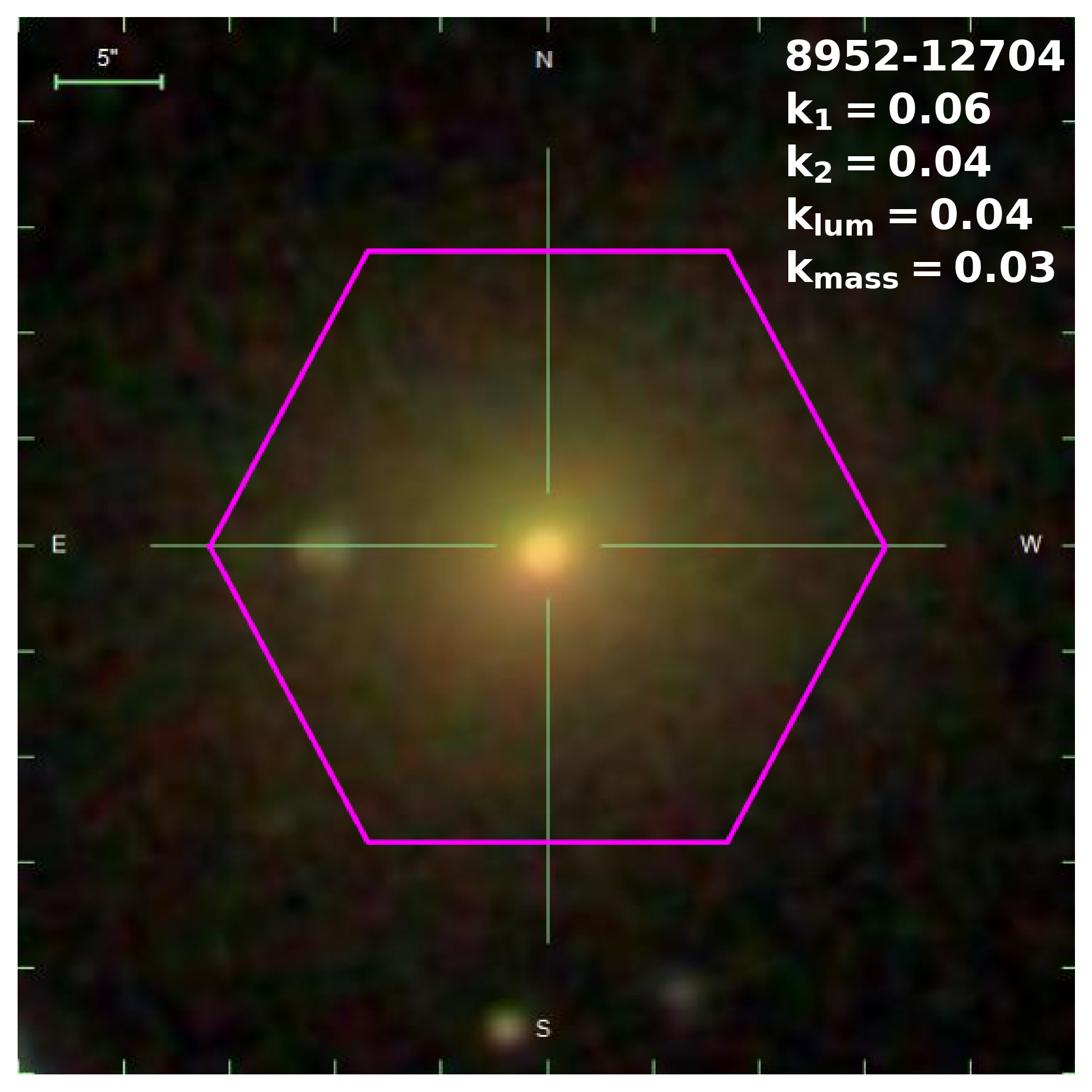}\hspace{0.015\textwidth}
    \includegraphics[scale=0.476,trim={0 1.3cm 0 0}]{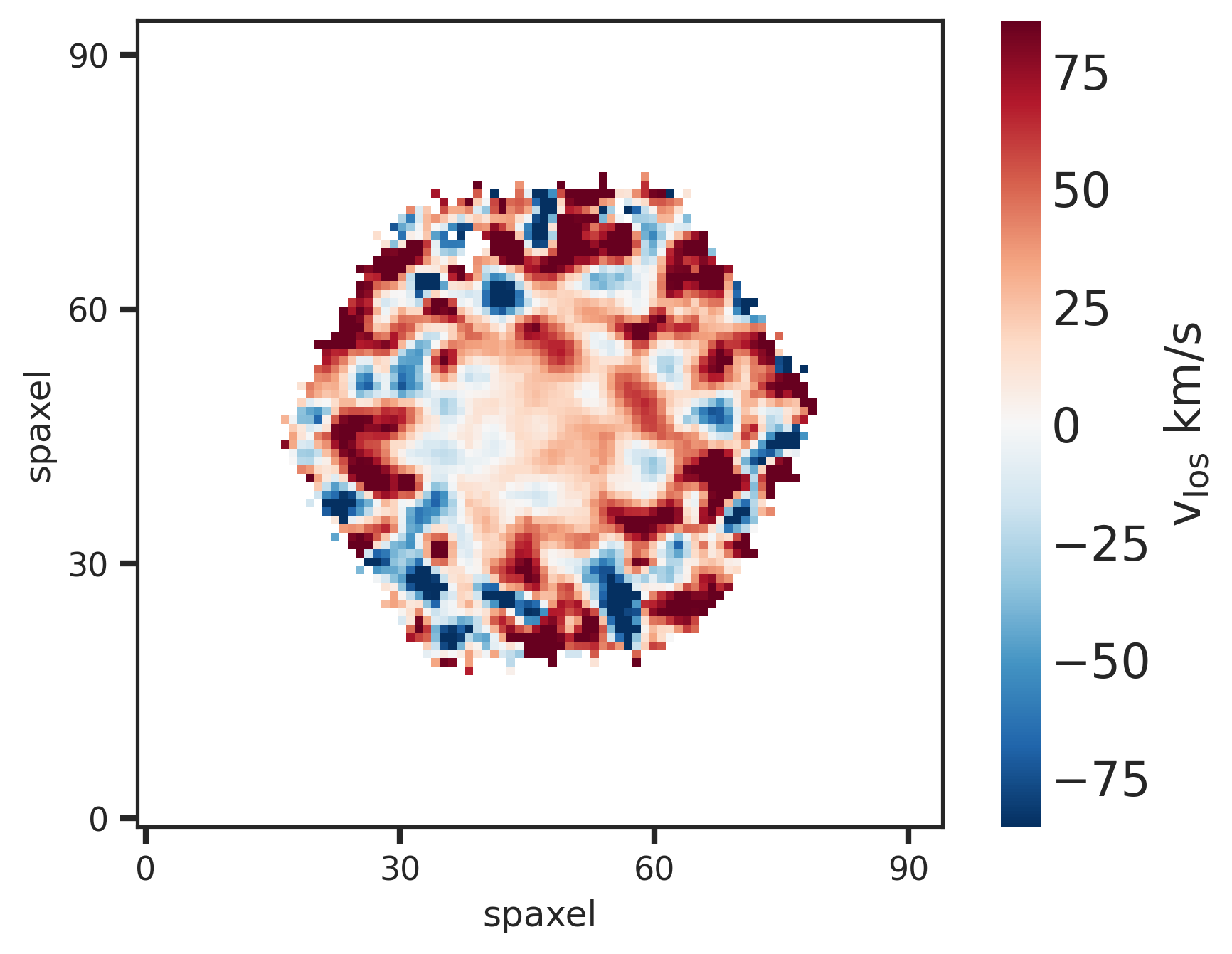}
    \includegraphics[scale=0.476,trim={0 1.3cm 0 0}]{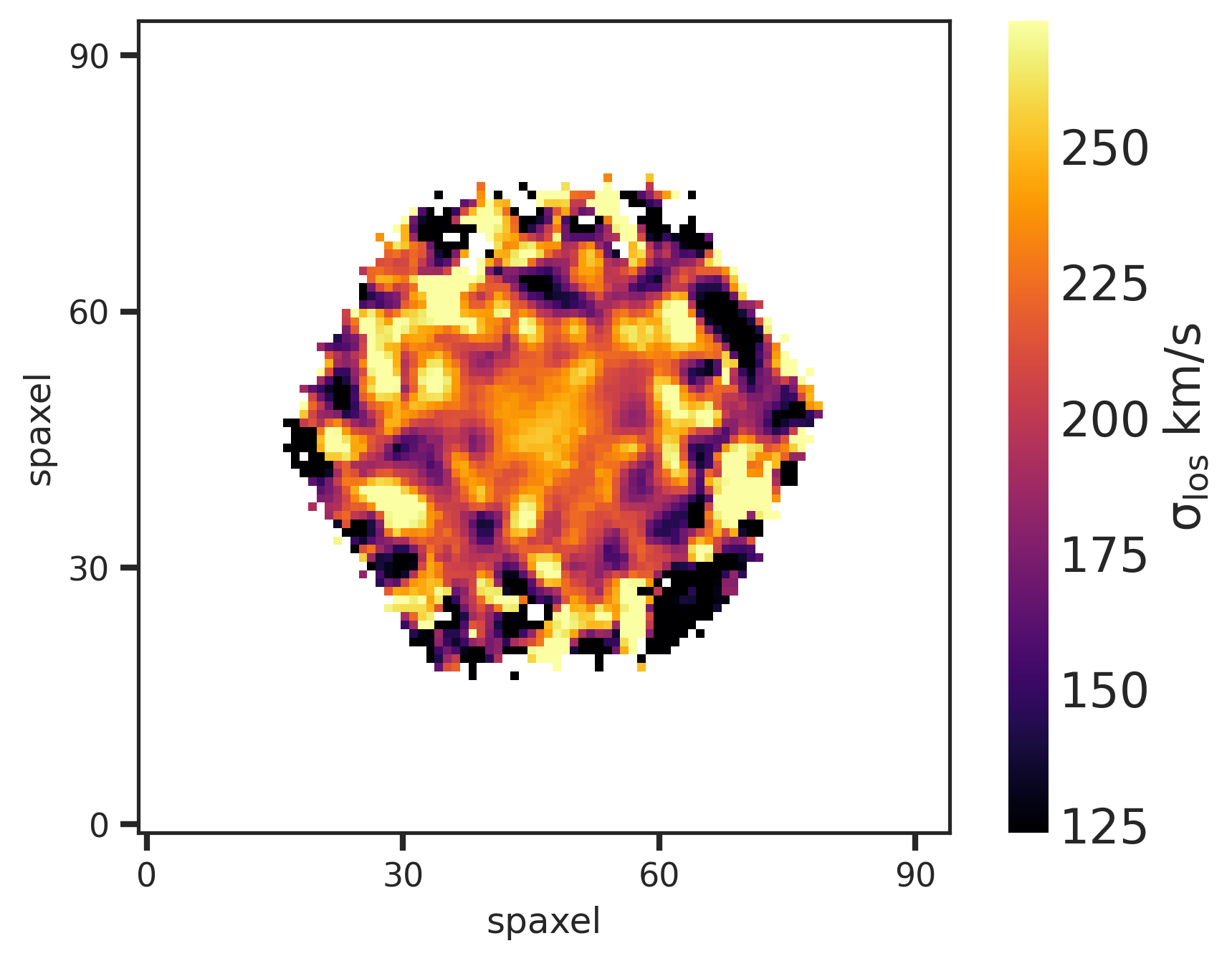}\vspace{0.02\textheight}\vspace{0.01\textheight}
    \includegraphics[scale=0.304]{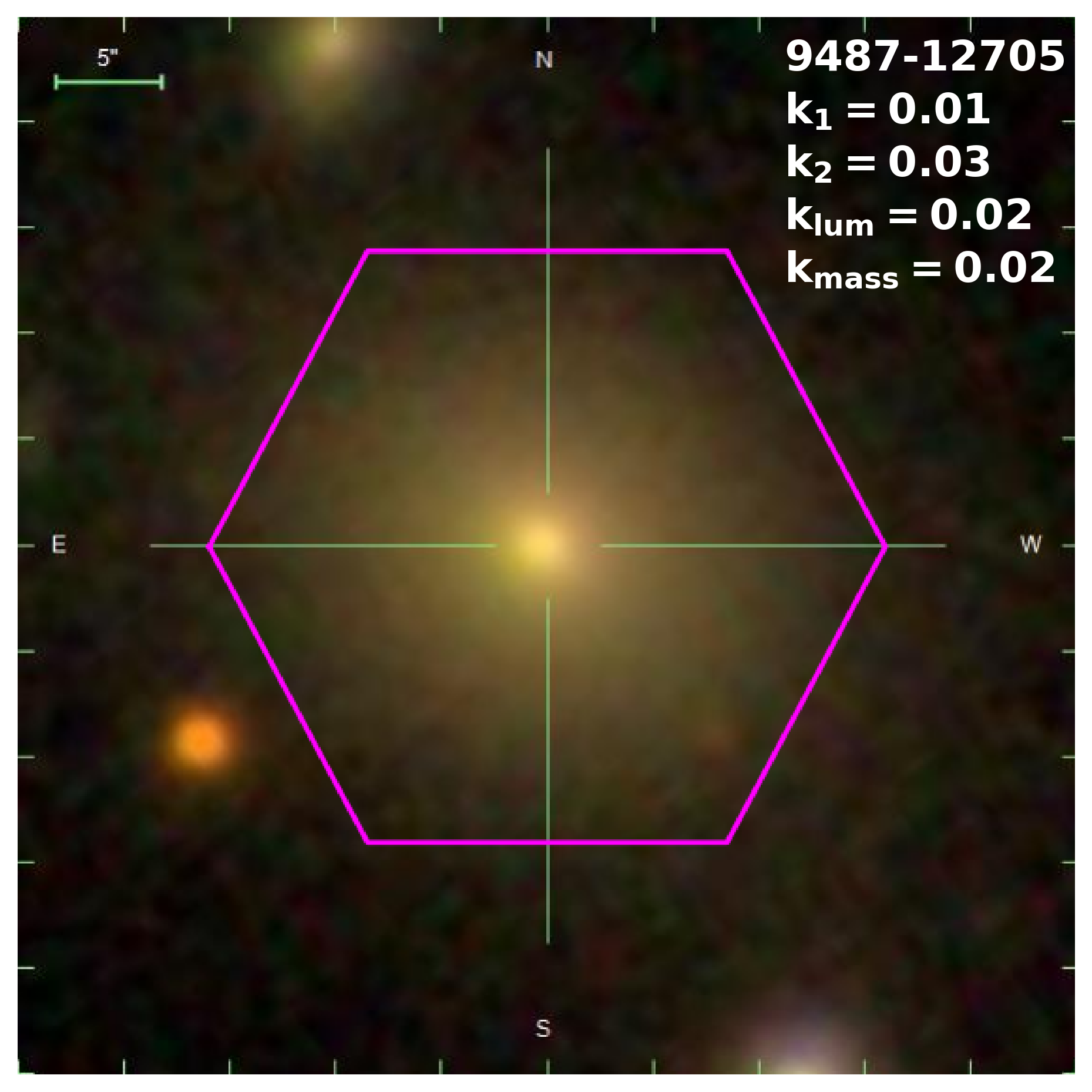}\hspace{0.015\textwidth}\includegraphics[scale=0.476,trim={0 1.3cm 0 0}]{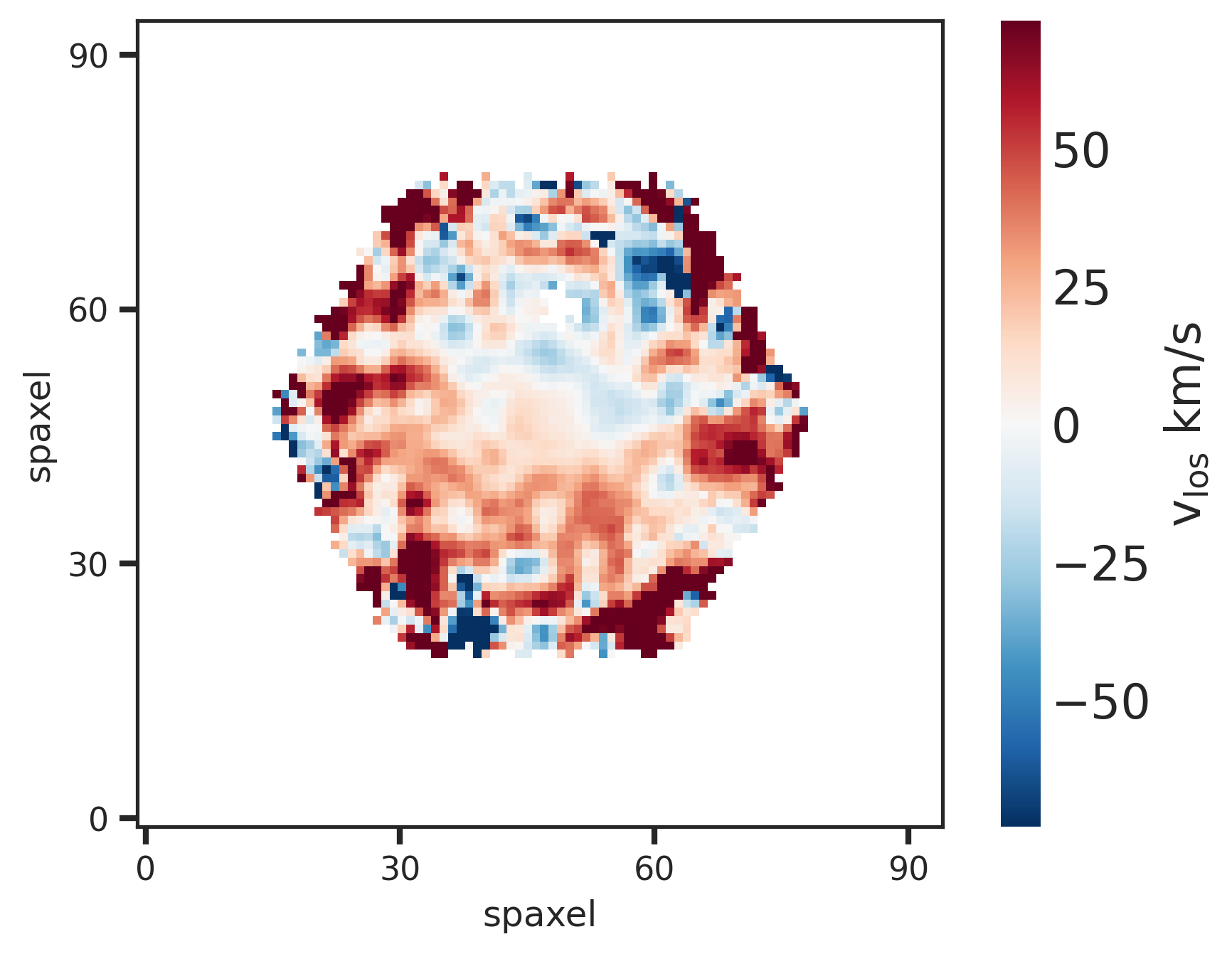}
    \includegraphics[scale=0.476,trim={0 1.3cm 0 0}]{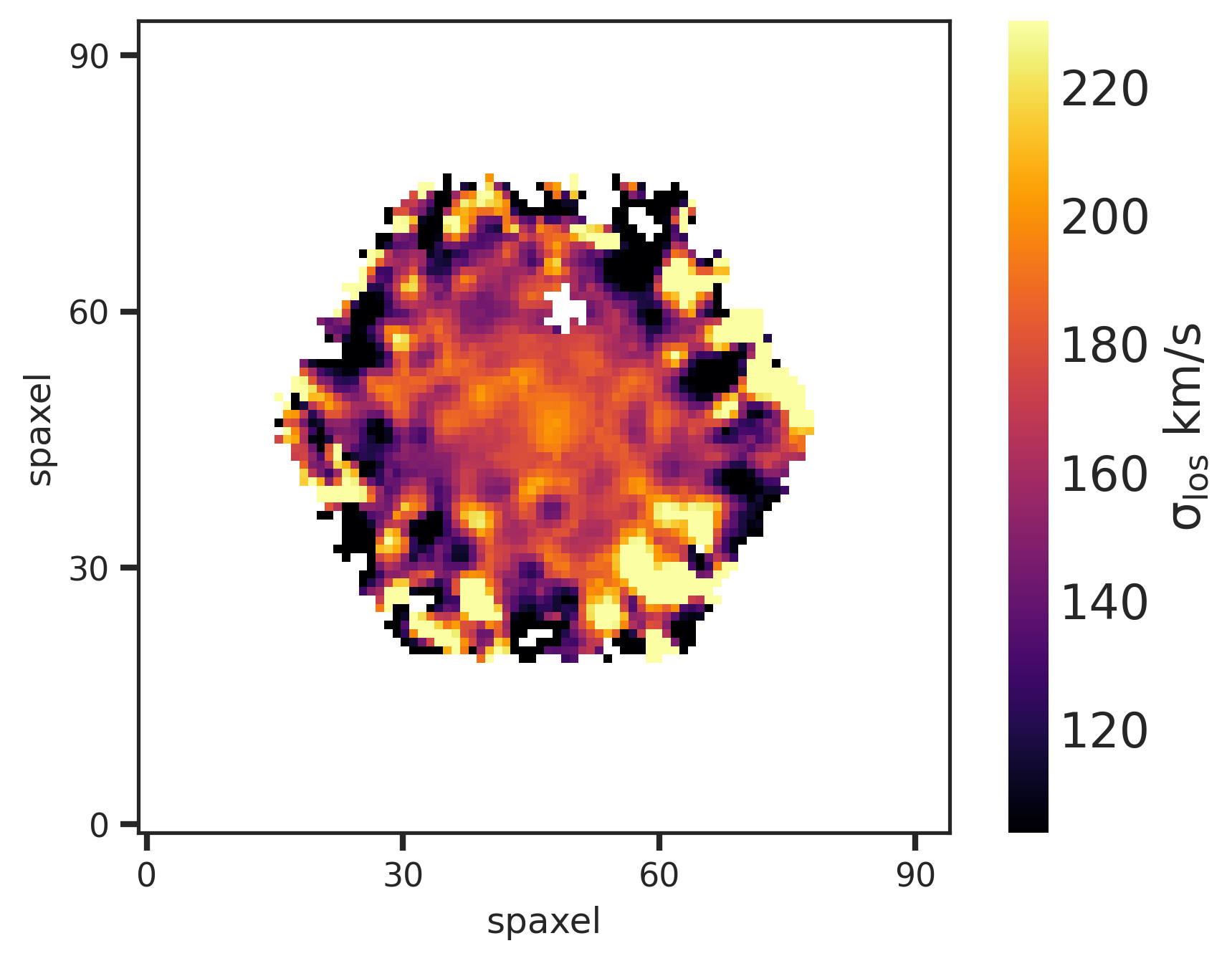}\vspace{0.02\textheight}
    \caption{Two examples of galaxies (i.e. 8952-12704 top row, 9487-12705 bottom row) with low $k_1$ and $k_2$. From left to right: SDSS \textit{gri} image, l.o.s. stellar velocity and l.o.s. stellar velocity dispersion. The magenta hexagon is the MaNGA field of view. }
    \label{fig:ex_small_k1_k2}
\end{figure*}

\subsection{Sample selection and Volume weights}
Similarly to what was done in Paper II we carefully selected, for the forthcoming analysis, a subset of $\simeq 5,000$ galaxies from the originally analysed MaNGA sample. Specifically, we discarded all galaxies with $\log_{10}{M_{\star}}<9.5$ as the stellar velocity dispersion is generally smaller than the instrumental velocity dispersion and those with bulge half-mass radius smaller (within its error) than the pixel size\footnote{The pixel size is $0.5''$ and the average full width at
half maximum (FWHM) of the SDSS i-band point spread function (PSF) is $\sim 1.2''$. Our choice is indeed similar to what done in \citep[][]{Gadotti_2009}{}. We refer to Paper II for a discussion about this.}. We also visually inspected and discarded all the objects showing evidence of ongoing mergers or other contaminants (i.e. wrong redshift, central non-masked stars and type I AGN). We cross-matched our sample with the catalogue presented in \cite{Fraser-McKelvie_2022} whose SFR measures have been collected from the GALEX-Sloan-WISE Legacy Catalogue 2 \citep[GSWLC-2][]{Salim_2016,Salim_2018}{}{}, obtaining a final sample of 4731 galaxies. 

The MaNGA stellar mass distribution is roughly flat above $10^9 ~ \rm M_{\sun}$ which means that galaxies at high stellar masses are oversampled compared to the galaxy stellar mass distribution observed in the local Universe. For this reason, a volume weight is assigned to each MaNGA galaxy \citep[][]{Wake_2017}{}{} to return the original sample to a volume-limited equivalent. Given that we limit our analysis to a subsample of the original MaNGA catalogue, we calculated corrected volume weights based on the fraction of galaxies in our sample with respect to the complete sample for each stellar mass and redshift bin (see \citealt{Fraser-McKelvie_2022} for further details).

\subsection{Model summary and parameter catalogue}
Throughout this paper we use the result of the analysis carried out in Paper II, we summarize in what follows the main assumption underlying our modelling strategy, we refer to Paper I and Paper II for a detailed discussion on the modelling approach. 

Each galaxy is assumed to be axially symmetric described as the superposition of a spherical bulge, two exponential
razor-thin discs and a dark-matter halo.
The galaxy position and orientation are characterized by the centre $(x_0,y_0)$, the position angle (P.A.) and the inclination $i$. Those are free parameters of the model and are estimated to provide the best projection of the galaxy onto the sky plane.

Each galaxy component is described through the following potential-density profiles:
\begin{itemize}
    \item the spherically symmetric and isotropic bulge is described by a Dehnen profile \citep{Dehnen1993} and parameterized in terms of its mass, scale radius and inner density slope (respectively $M_b$, $R_b$ and $\gamma$).
    The volume density and mean velocity dispersion of the bulge are projected to provide the bulge surface density and line-of-sight (l.o.s.) velocity dispersion. The surface density is then converted into surface brightness through the bulge mass-to-light ratio ($M/L_b$). Throughout the paper, we refer to this component as "dispersion-supported bulge" because it is identified as a non-rotating, isotropic and spherically symmetric component;

    \item two (i.e. "inner" and "outer") axially symmetric components are described by exponential razor-thin profiles characterized in terms of their masses ($M_{d,1}$, $M_{d,2}$), scale radii ($R_{d,1}$, $R_{d,2}$) and two further parameters for describing their kinematical state ($k_1$, $k_2$). More precisely, we approximate the intrinsic velocity dispersion and the tangential velocity of the inner (outer) disc assuming that a fraction $k_1$ ($k_2$) of the total kinetic energy is in ordered bulk motion and the rest goes into an isotropic velocity dispersion component. The discs surface density is converted into surface brightness through the disc mass-to-light ratios ($M/L_{\rm d,1}$, $M/L_{\rm d,2}$). All these quantities are then projected along the l.o.s.. Note that, the "kinematic decomposition parameters" ($k_1$ and $k_2$) are free parameters of the model whose value ranges from $0$ for a dispersion-supported regime to $1$ for a rotation-supported regime. The freedom given to the inner and the outer disc to be either dispersion-supported or rotationally-supported allows us to model a variety of structures present in galaxies resulting in multiple possible interpretations of the inner and the outer "disc natures" (see Sec.~\ref{sec:Global and Structural kinematics} for a detailed discussion);
    
    \item the Navarro-Frenk-White \citep{Navarro1996} dark matter halo parameterized in terms of the halo-to-stellar mass fraction and the halo concentration (respectively $f_\star$ and $\rm c$). It is the only dark component in our model contributing to the overall potential. Note that, given the limited spatial extension of the MaNGA data, only the mass within the observed kinematics can be robustly recovered. Indeed, as a more conservative approach, for this analysis, we considered the parameters estimated in the "halo-fixed" configuration (see Paper II for further details) ;

\end{itemize}

The total surface brightness of a galaxy model is the sum of the surface brightness of each visible component. The total l.o.s. velocity and l.o.s. velocity dispersion are determined as a brightness-weighted average of each component contributions\footnote{Note the by definition the bulge l.o.s. velocity is zero} while the average mass-to-light ratio is defined as the fraction between the total surface density and the total surface brightness. Note that since every component depends on its own mass-to-light ratio the average mass-to-light ratio will not be radially constant.  

We refer to Tab.~\ref{tab:parameters} for a summary of all the model parameters together with a brief description. All the free parameters of the model are determined by fitting simultaneously the brightness, kinematics and mass-to-light ratio data with a nested sampling algorithm \citep[][]{Skilling2004}{}{}. The model evaluation is performed on GPUs. For each galaxy, models with a different number of visible components are compared with each other. The best model is chosen to be the one with the highest Bayesian evidence.



\section{Kinematics of the whole galaxies and of their components}
\label{sec:Global and Structural kinematics}

\begin{figure*}
    \includegraphics[scale=0.304]{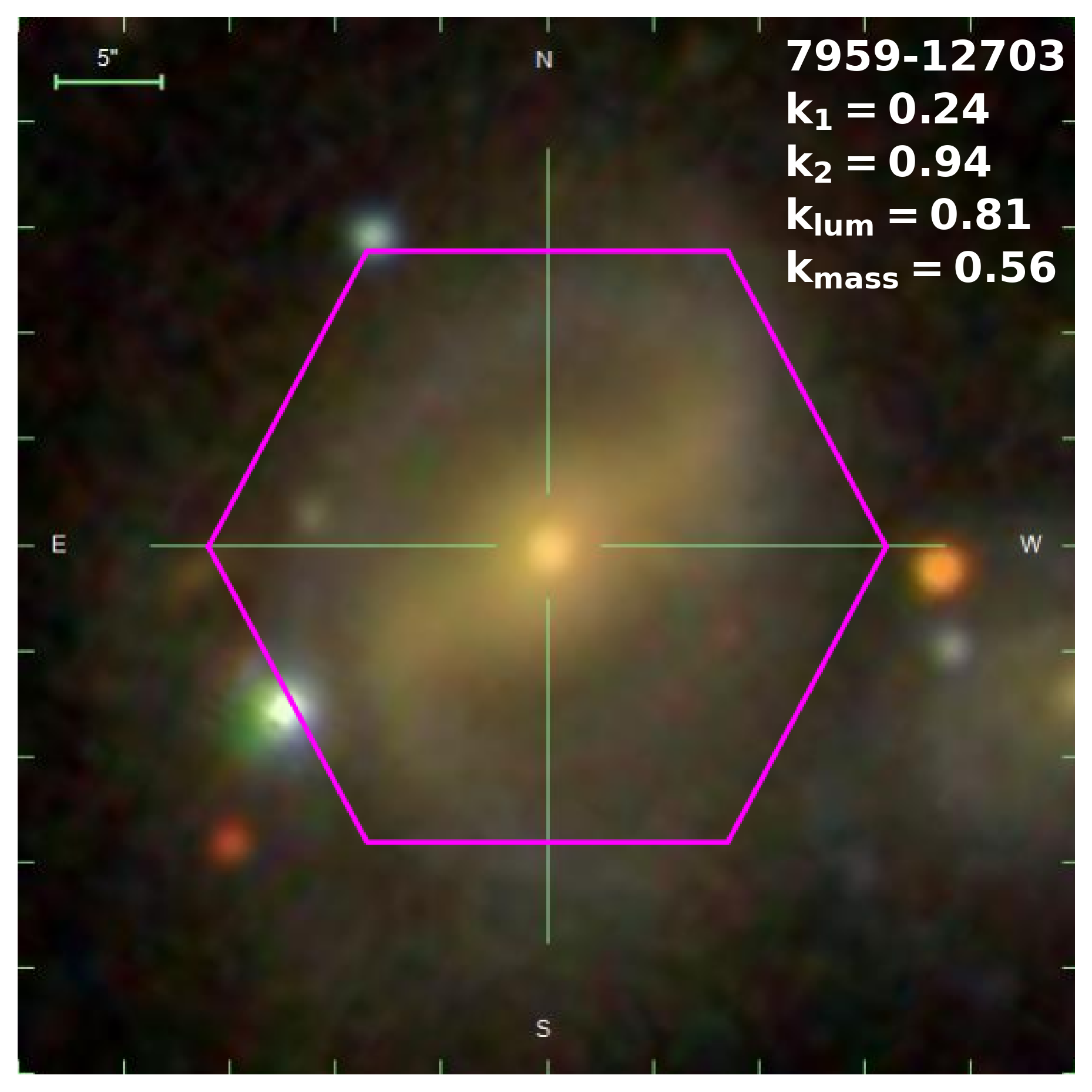}\hspace{0.015\textwidth}\includegraphics[scale=0.476,trim={0 1.3cm 0 0}]{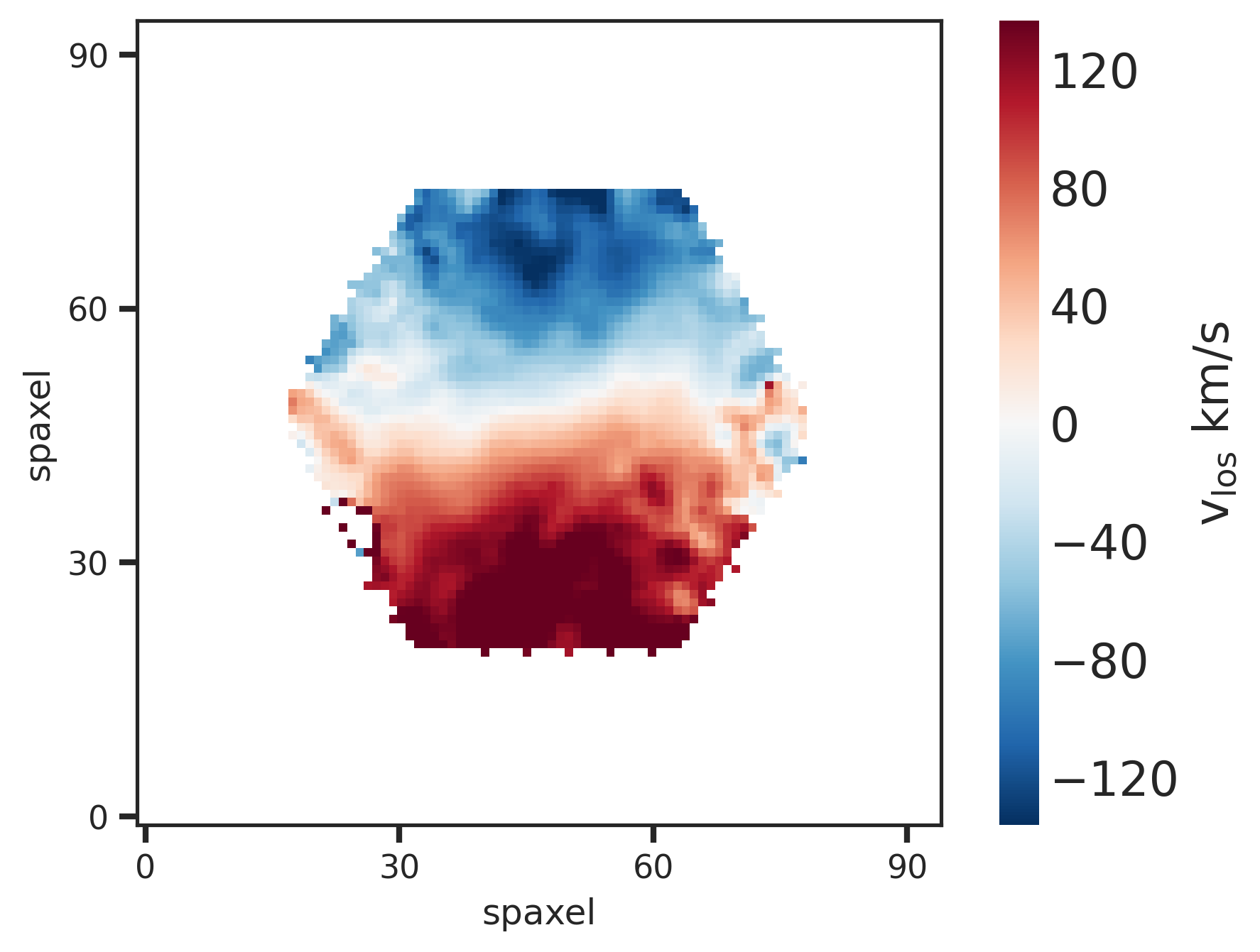}
    \includegraphics[scale=0.476,trim={0 1.3cm 0 0}]{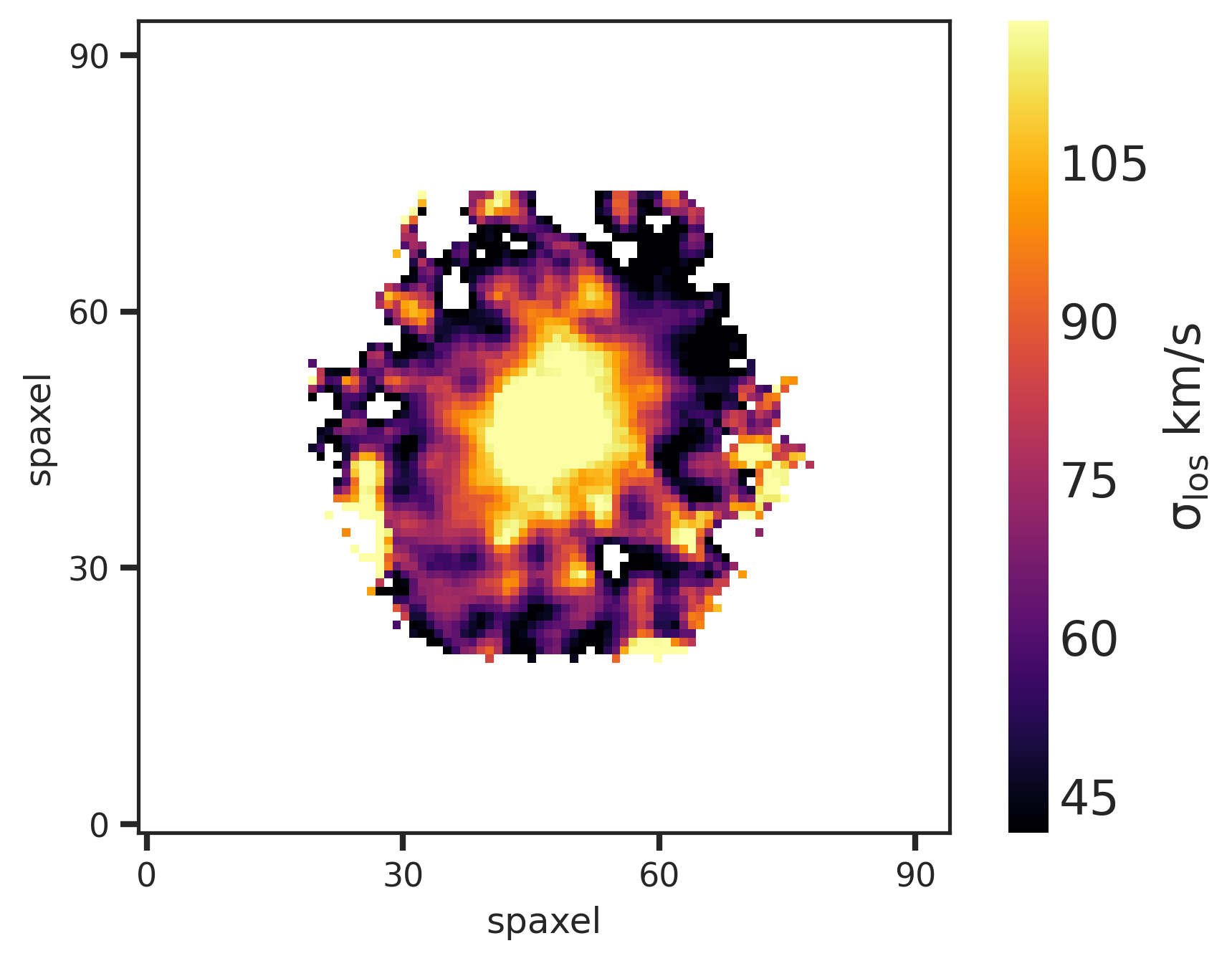}\vspace{0.02\textheight}\vspace{0.01\textheight}
    \includegraphics[scale=0.304]{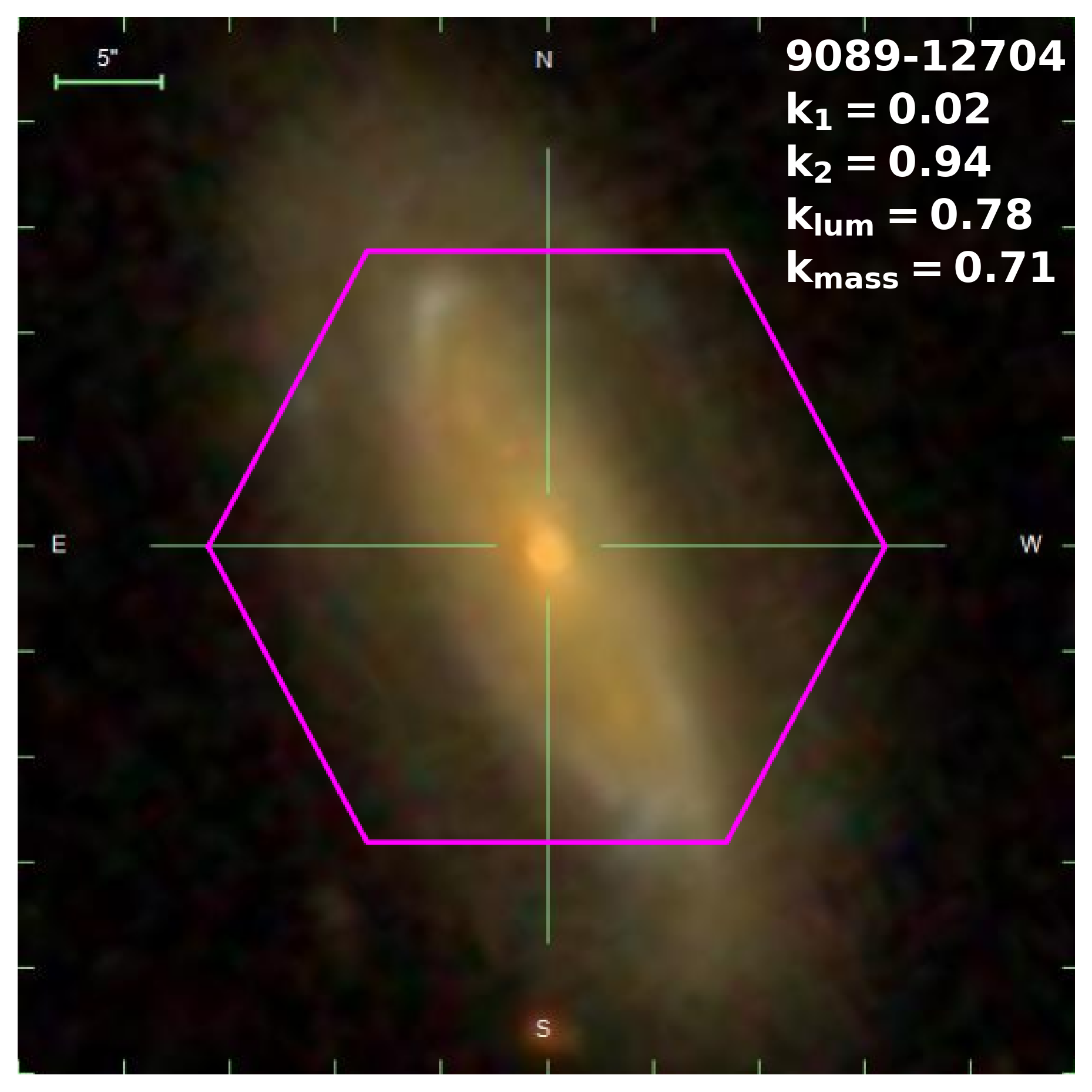}\hspace{0.015\textwidth}\includegraphics[scale=0.476,trim={0 1.3cm 0 0}]{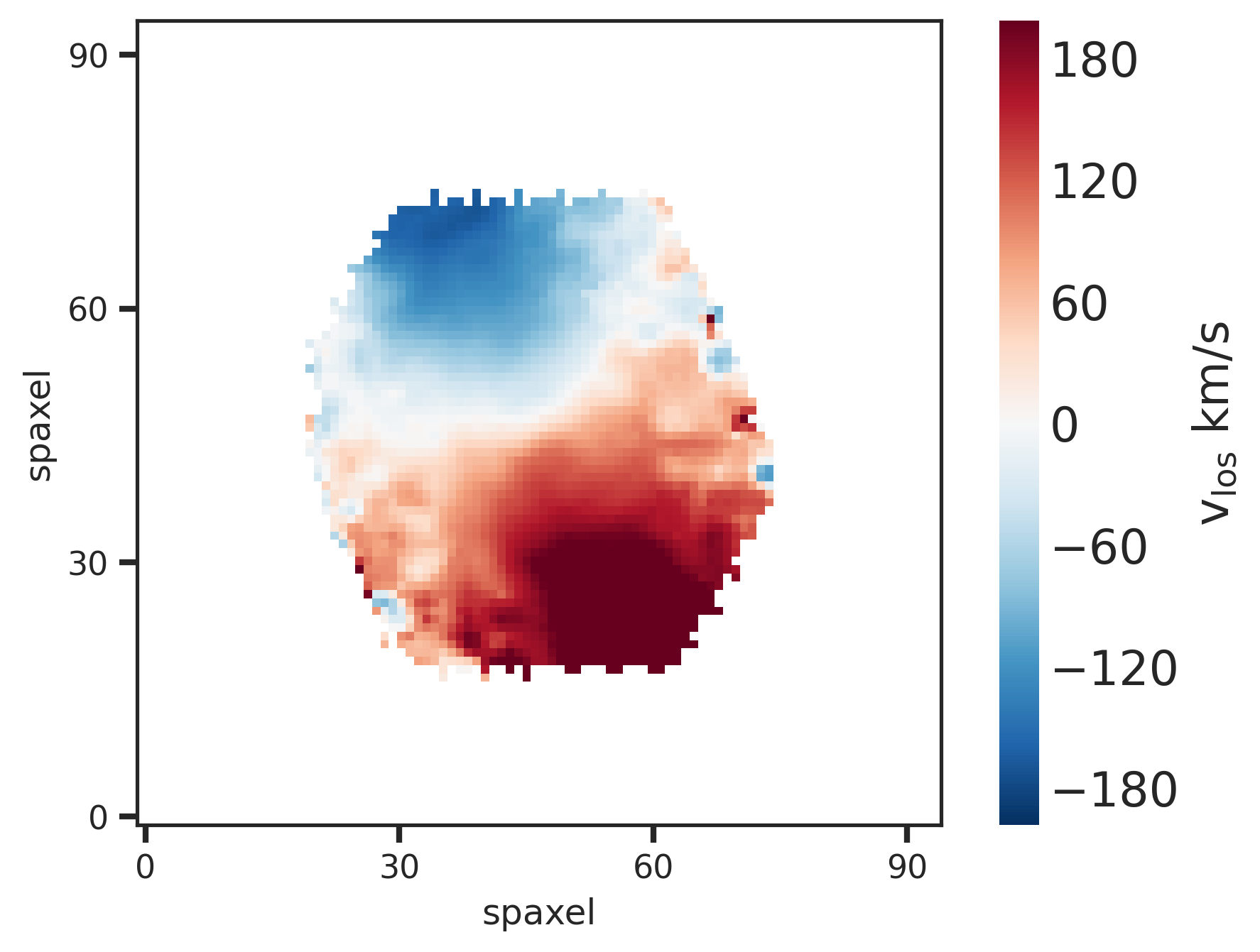}
    \includegraphics[scale=0.476,trim={0 1.3cm 0 0}]{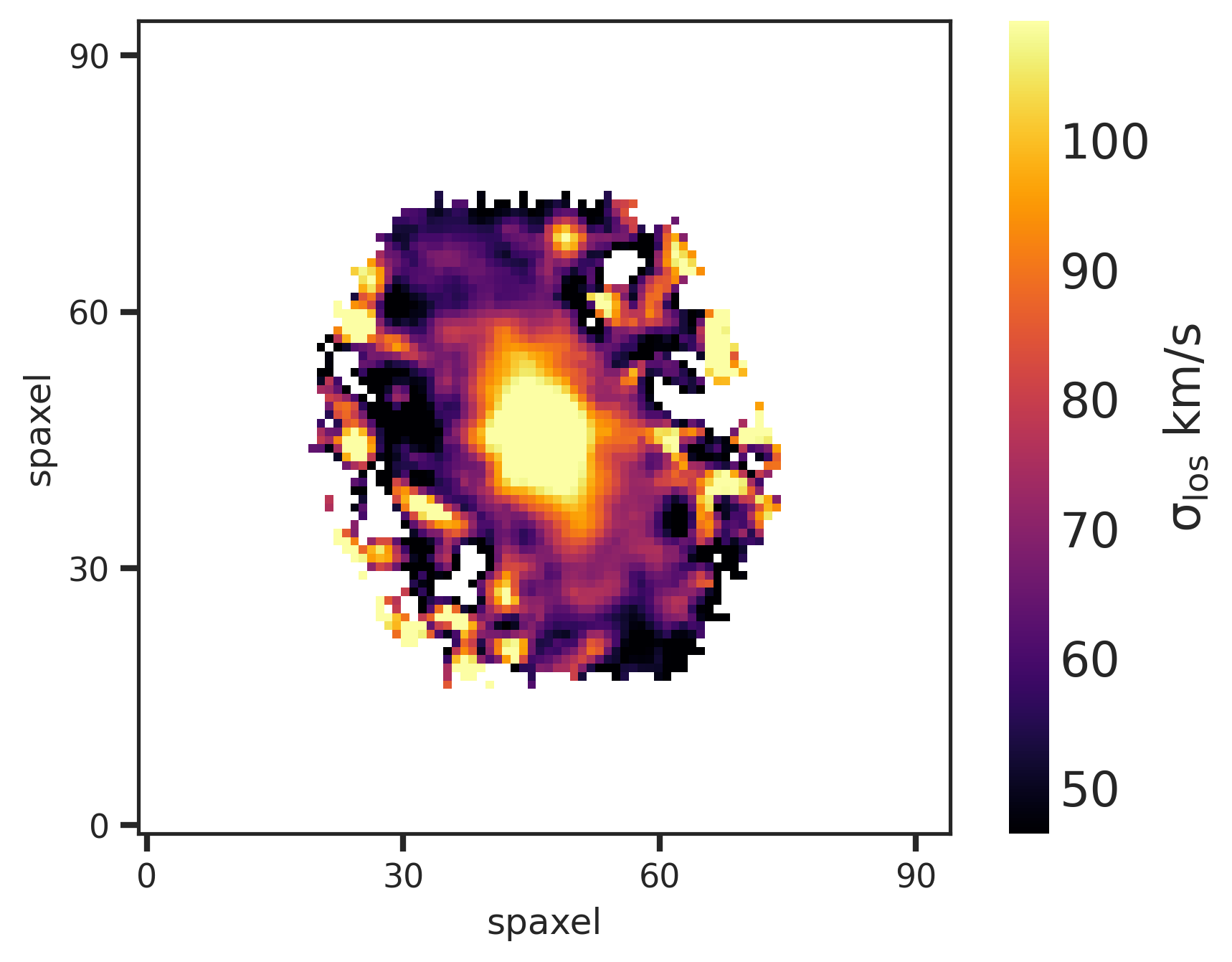}\vspace{0.02\textheight}
    \caption{Two examples of galaxies (i.e. 7959-12703 top row, 9089-12704 bottom row) with low $k_1$ and high $k_2$. From left to right: SDSS \textit{gri} image, l.o.s. stellar velocity and l.o.s. stellar velocity dispersion. The magenta hexagon is the MaNGA field of view.}
    \label{fig:ex_small_k1_high_k2}
\end{figure*}

In this section, we aim to clarify the meaning of the $k_1$ and $k_2$ parameters and to define new parameters of crucial relevance in the forthcoming analysis.


The left and middle panels of Fig.~\ref{fig:k_distributions} show the distribution of the $k_1$ and the $k_2$ parameters in our sample.
Focusing on the left panel we see a clear peak in the histogram for $k_1 \lesssim 0.1$ while the distribution is almost flat for higher values. When $k_1$ is close to zero the inner disc models a dispersion component in the galaxy centre which may be exponential in its light profile or, more often, not axi-symmetric when projected along the l.o.s.; in these cases, the inner disc is providing a correction to the dispersion-supported bulge. We note that a non-negligible amount of galaxies have intermediate values of $k_1$. This is not surprising since structures such as thick discs, pseudobulges and bars are common in disc-like galaxies resulting in galaxies with intermediate rotational support \citep[][]{Fraser-McKelvie_2022}{}{}. Unfortunately, given the simplicity of our current model, it is still not possible to clearly distinguish among these structures.
The middle panel of Fig.~\ref{fig:k_distributions} shows a bimodality in the distribution of $k_2$; this reflects, to some extent, the bimodality between rotating discs and pressure-supported ellipticals. When $k_2$ is close to one ($\gtrsim 0.6$) the outer disc models the large-scale rotating disc of the galaxy; on the contrary, when $k_2$ is close to zero ($\lesssim 0.10$), in the majority of the cases (we discuss some exceptions in Appendix~\ref{app:other_k1_k2_examples}), the galaxy is a fully dispersion-supported system (i.e., a genuine elliptical). The right panel of Fig.~\ref{fig:k_distributions} shows the distribution of the $k_{\rm lum}$ (in turquoise) and $k_{\rm mass}$ (in dark green) which are the luminosity-weighted and the mass-weighted average of the $k_1$ and $k_2$ parameters respectively defined as:

\begin{equation}
    \label{eq:k_lum}
    k_{\rm lum} = \frac{k_1 L_{\rm d,1}+k_2 L_{\rm d,2}}{L_{\rm b}+L_{\rm d,1}+L_{\rm d,2}},
\end{equation}

\begin{equation}
    \label{eq:k_mass}
    k_{\rm mass} = \frac{k_1 M_{\rm d,1}+k_2 M_{\rm d,2}}{M_{\rm b}+M_{\rm d,1}+M_{\rm d,2}},
\end{equation}

\begin{figure*}
    \includegraphics[scale=0.304]{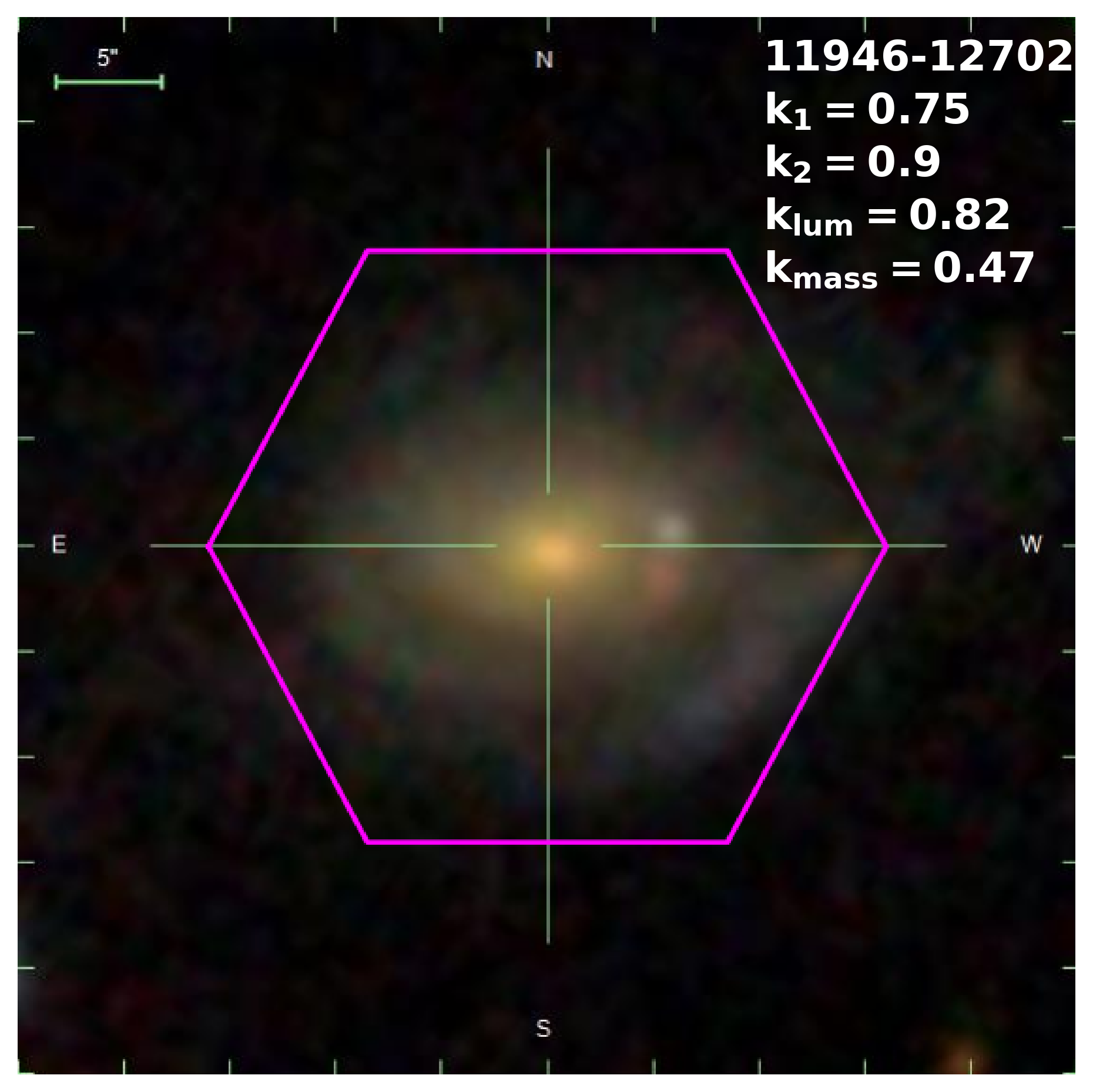}\hspace{0.015\textwidth}\includegraphics[scale=0.476,trim={0 1.3cm 0 0}]{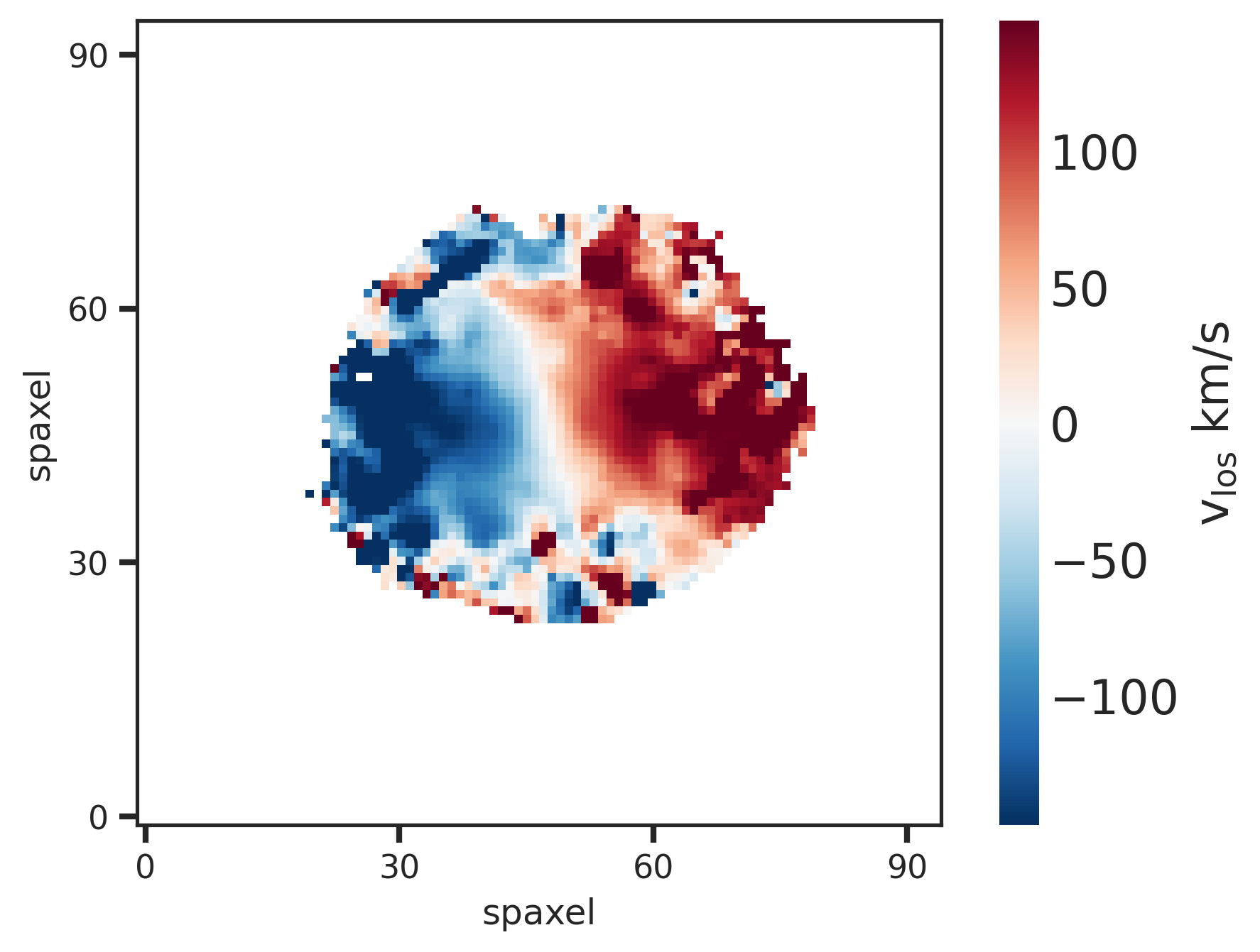}
    \includegraphics[scale=0.476,trim={0 1.3cm 0 0}]{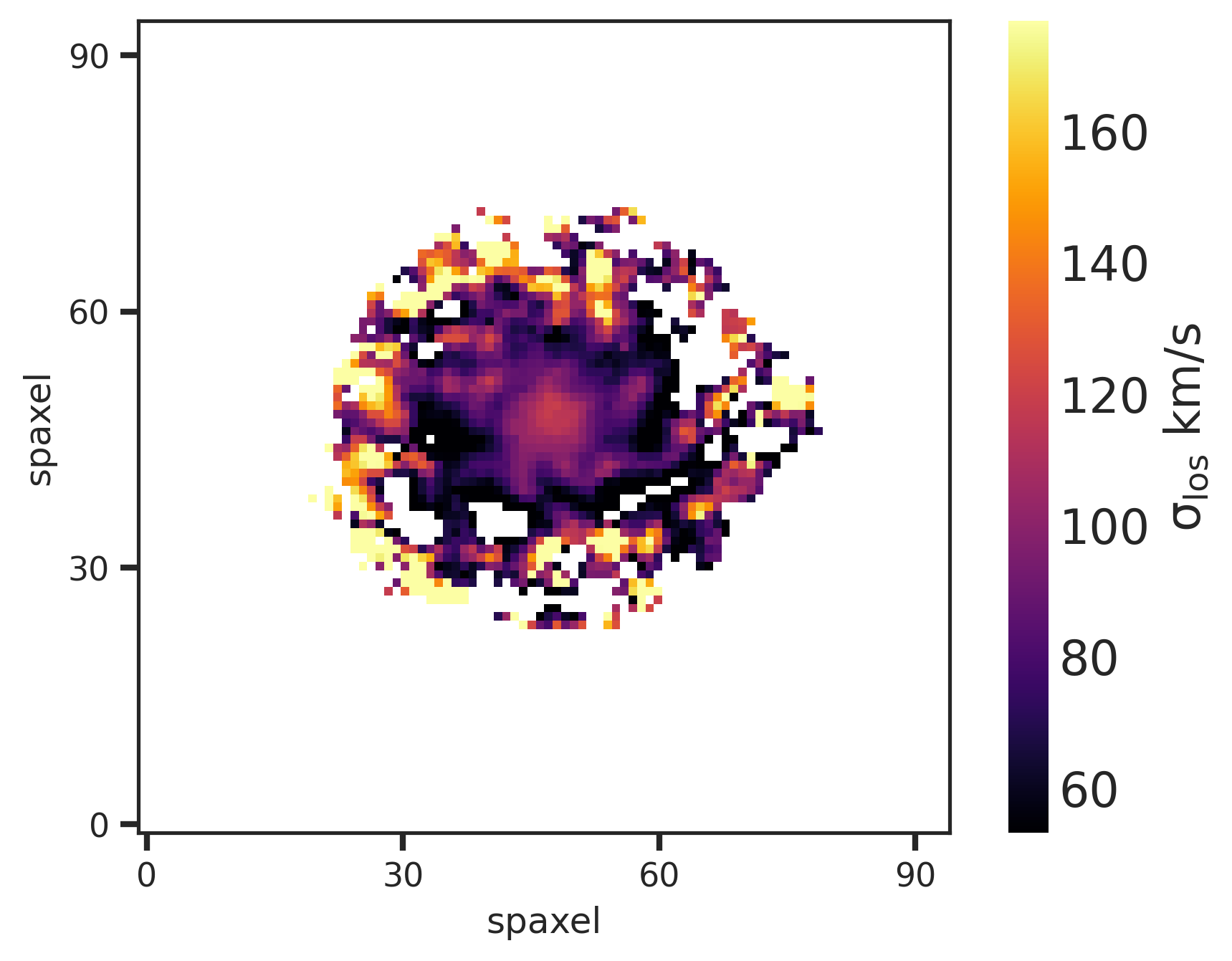}\vspace{0.02\textheight}\vspace{0.01\textheight}
    \includegraphics[scale=0.304]{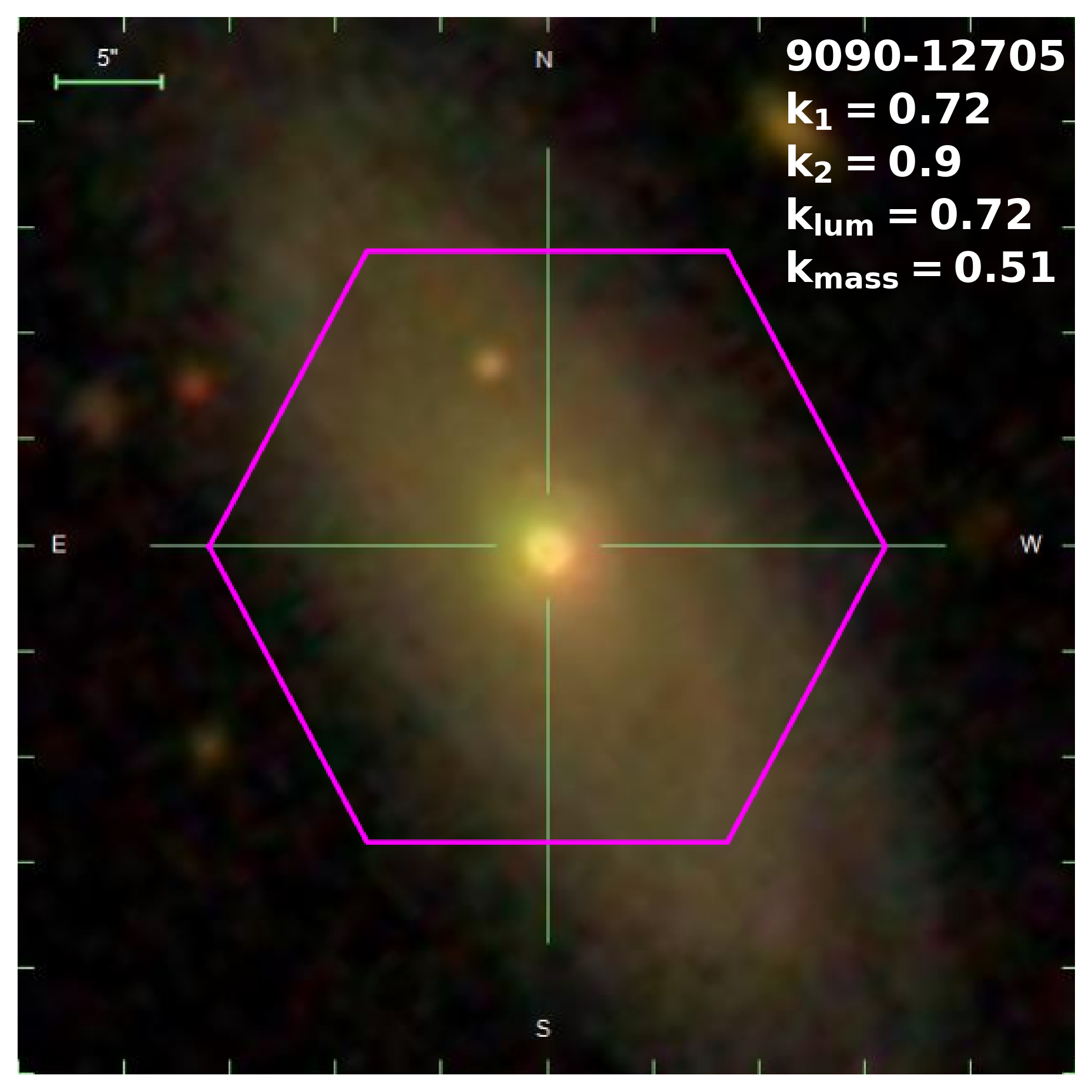}\hspace{0.015\textwidth}\includegraphics[scale=0.476,trim={0 1.3cm 0 0}]{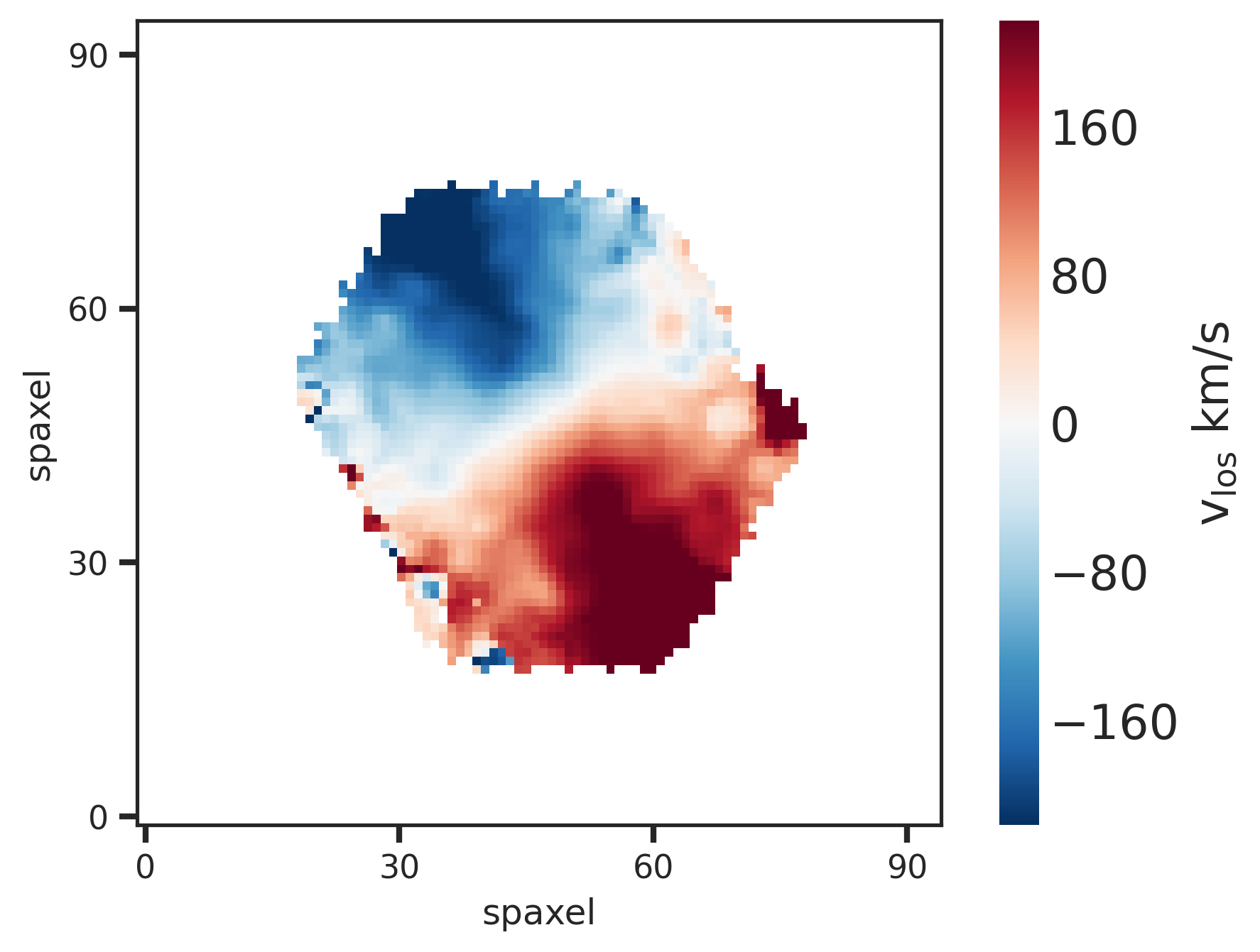}
    \includegraphics[scale=0.476,trim={0 1.3cm 0 0}]{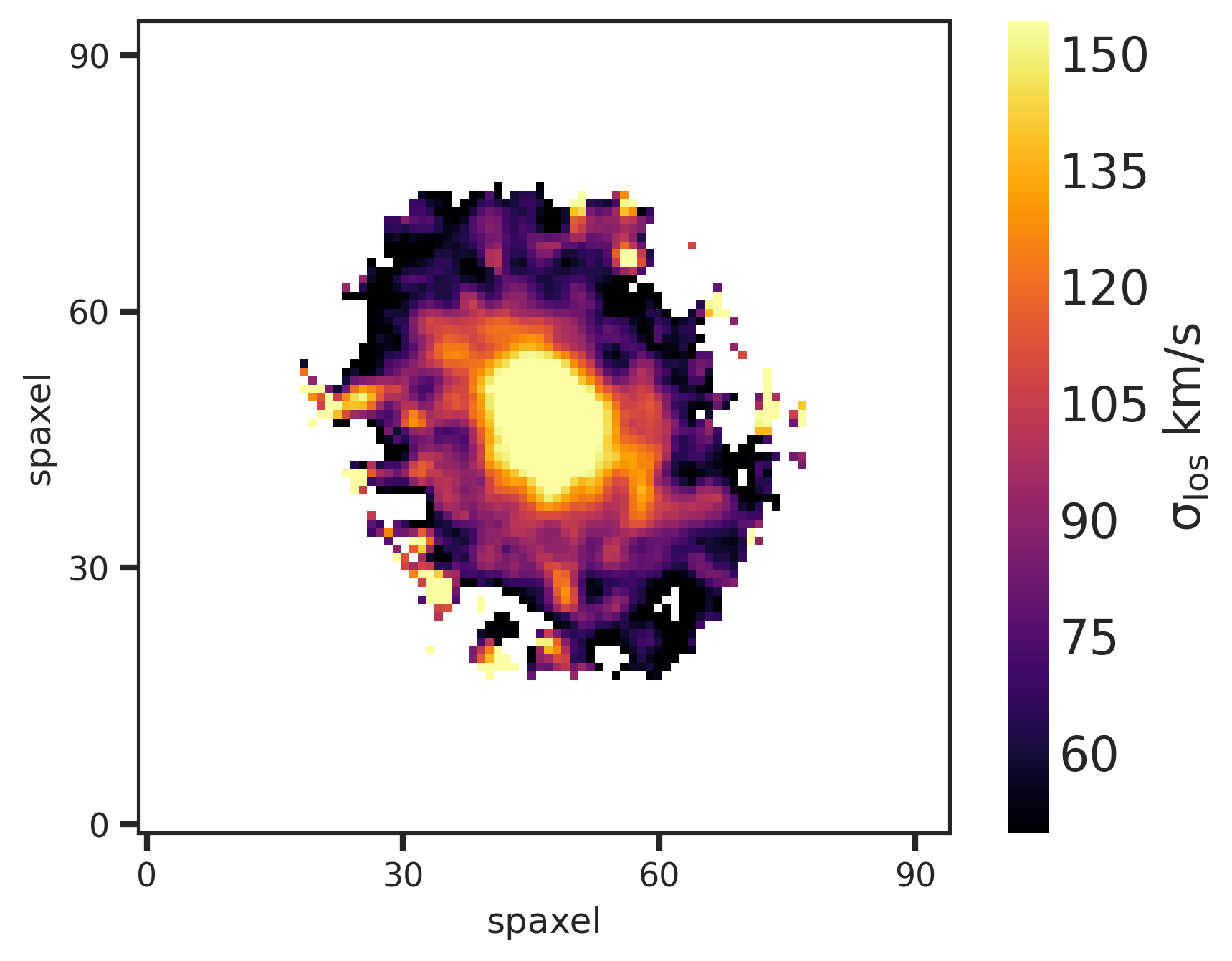}\vspace{0.02\textheight}
    
    \caption{Two examples of galaxies (i.e. $11946-12702$ top row, $9090-12705$ bottom row) with high $k_1$ and high $k_2$. From left to right: SDSS \textit{gri} image, l.o.s. stellar velocity and l.o.s. stellar velocity dispersion. The magenta hexagon is the MaNGA field of view.}
    \label{fig:ex_high_k1_high_k2}
    
\end{figure*}

\noindent where $L_j = M_j/(M/L)_j$ with $j=\{\mathrm{b;d,1; d,2}\}$ is the luminosity of the j-th component obtained dividing the mass $M_j$  of the j-th component by its mass-to-light ratio ($(M/L)_j$)\footnote{For each luminous o baryonic component we consider the mass within $R_{\rm max}$ defined as the maximum radius along the galaxy major axis for which MaNGA provides the kinematic. This choice is similar to what was done in Paper II; we verified that the total stellar mass within $R_{\rm max}$ is consistent with the measurements provided by the NASA-Sloan Atlas (NSA) \url{http://nsatlas.org/}.}. $k_{\rm lum}$ ($k_{\rm mass}$), being the luminosity (mass) -weighted average of $k_1$ and $k_2$ among all the components, quantifies the dynamical state of the galaxy as a whole\footnote{We refer to Appendix \ref{app:comparison_with_lambdaR} and to Paper II for a comparison of $k_{\rm lum}$ with the more commonly adopted kinematic tracers $\lambda_R$ and $V/\sigma$.}. It is important to note that the meaning of this parameter is different from $k_1$ and $k_2$. While $k_1$ and $k_2$ are related to the kinematics of individual galaxy components, preserving information about the galaxy structure, $k_{\rm lum}$ and $k_{\rm mass}$ characterize the overall kinematic state of the entire galaxy. Therefore, it is not unexpected to encounter galaxies with different $k_1$ or $k_2$ parameters and yet similar $k_{\rm lum}$ or  $k_{\rm mass}$, due to the differing relative importance of each galaxy component.  
The distribution of $k_{\rm lum}$ (third panel, turquoise colour in Fig.~\ref{fig:k_distributions}) shows a clear peak close to $k_{\rm lum} \simeq 0.75$ corresponding to fast-rotating galaxies with a flattening toward $k_{\rm lum} \simeq 0$. Interestingly, the peak at zero, present both in the $k_1$ and $k_2$ distributions, disappears. This is because the majority of the galaxies with an inner dispersion-dominated component (i.e. $k_1\simeq0$) when analysed as a whole, demonstrate clear rotation, with the outer disc retaining most of the angular momentum of the galaxy and harbouring a significant fraction of the mass/light of the system. Slow-rotating galaxies (i.e. $k_1 \simeq 0$ and $k_2 \simeq 0$) are still present but, statistically, their number contribution is less relevant. The same behaviour is seen for $k_{\rm mass}$, albeit with a broader distribution and a peak shifted towards smaller values. This can be primarily attributed to the definition of $k_{\rm mass}$, where the contribution of rotationally-dominated systems, particularly the outer one, holds more weight in luminosity than it does in mass, especially in star-forming galaxies. 

For the sake of clarity, we provide some examples showing different behaviour of the $k_1$ and $k_2$ parameters. We roughly identified three main groups in the $(k_1; k_2)$ space, namely\footnote{Exceptions to these three classes are present even though they do not play a relevant role in our discussion; for completeness, we provide some show-case examples in Appendix~\ref{app:other_k1_k2_examples}.}:

\begin{itemize}
    \item galaxies with low $k_1$ and low $k_2$.
    
    Fig.~\ref{fig:ex_small_k1_k2} shows two examples (plate-IFUs: 8952-12704 and 9487-12705) of galaxies with $k_1$, $k_2$ and also $k_{\rm lum}$ and $k_{\rm mass}$ close to zero. These two galaxies, according to the MaNGA Visual Morphology Catalogue \citep[VMC-VAC,][]{Vazquez_2022}{}{}, are classified as ellipticals and they do not show any clear rotational pattern. This should be enough to confidently state that no rotationally dominated component is present in these objects. Despite that, almost by construction, our best fit for these galaxies comprises two, additional components, the inner and the outer disc. However, these are needed primarily to compensate for our assumption that the dispersion-supported central component is purely spherical, while in this case the galaxy clearly shows a significant ellipticity. Nevertheless, from all points of view, both objects are dispersion-dominated. 
    
    \item galaxies with low $k_1$ and high $k_2$.
    
    Fig.~\ref{fig:ex_small_k1_high_k2} shows two examples (plate-IFUs: 7959-12703 and 9089-12704) of galaxies with low $k_1$ and high $k_2$. As can be seen in the figure, both galaxies show features in their light profiles (e.g. bars, spirals and rings), and they have a clear rotational pattern with a quite significant central peak in the velocity dispersion. These galaxies have $k_{\rm lum}$ above 0.75, suggesting, as expected, the dominance of a disc. In both cases, the inner disc models the central, dispersion-dominated region of the galaxy. It is hard to interpret them as thin discs; more likely, they provide a correction to the inner region of the galaxies where the possibly secular bulge, given the presence of spirals and rings, may deviate from spherical symmetry. 7959-12703 has a higher $k_1$ and a smaller bulge-to-inner-disc ratio (i.e. $L_b/L_{\rm d1}$) compared to 9089-12704, looking at the data, it seems that 7959-12703 has slightly larger rotational support and a more disc-like light distribution in the centre.
    
    \item galaxies with high $k_1$ and high $k_2$.
    
    Fig.~\ref{fig:ex_high_k1_high_k2} shows two examples (plate-IFUs: 11946-12702 and 9090-12705) of galaxies with high $k_1$, $k_2$.  
    Both galaxies show quite high l.o.s. velocity (middle panels of Fig.~\ref{fig:ex_high_k1_high_k2}) even near the centre, explaining why a rotationally-supported inner disc is required by the model. In these cases,  the inner discs are spatially probing the transition region (i.e. between the bulge and the disc) of the galaxy, mostly contributing to the dynamically cold component of the galaxy (i.e. the galactic disc).
\end{itemize}
\begin{figure}
    \centering
    \includegraphics[scale=0.6]{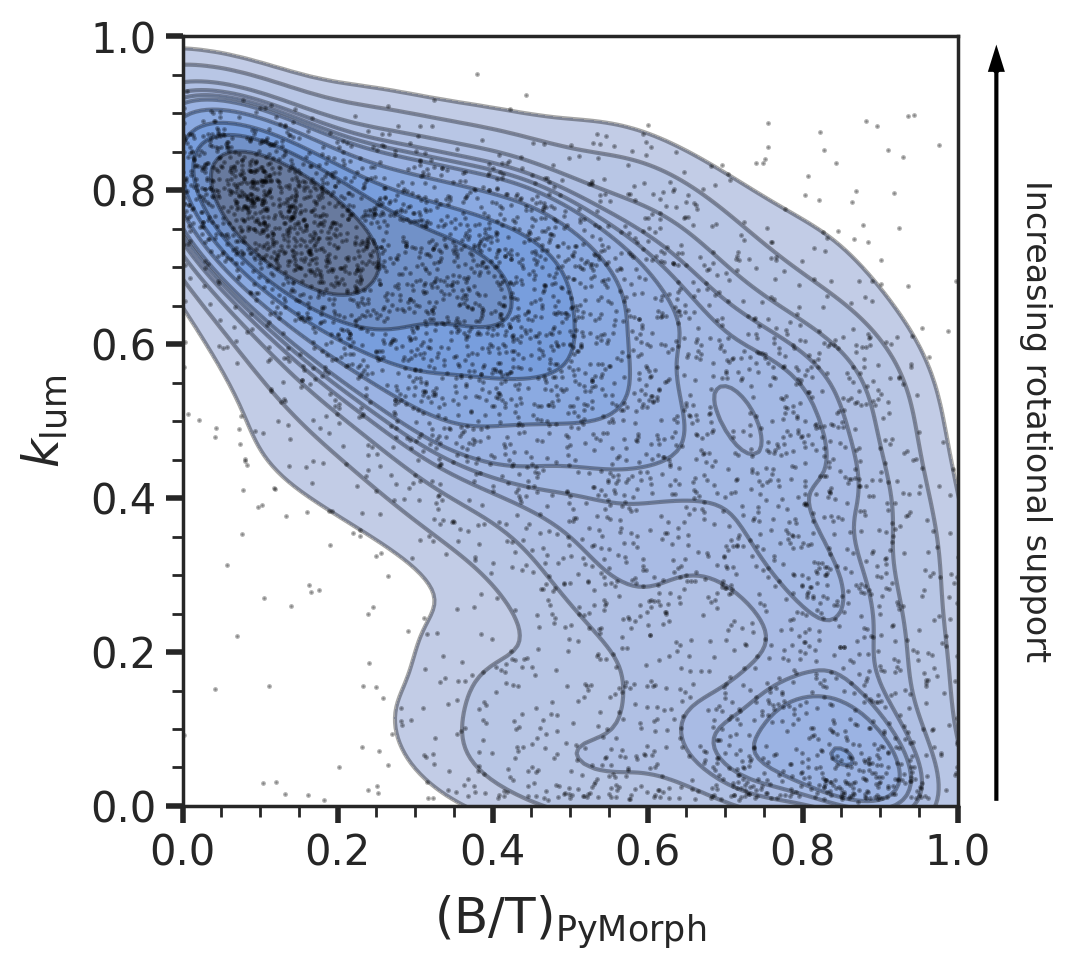}
    \caption{Comparison between the luminosity-weighted kinematic tracer $k_{\rm{lum}}$ (vertical axis) and a purely photometric bulge-to-total ratio taken from \cite{Sanchez2022} (horizontal axis). The contours (grey lines filled with the blue shaded colours) are drawn at probability levels with a constant spacing of 0.1 while the black dots represent the data. \fr{The vertical arrow on the right side qualitatively explains the meaning of the $k_{\rm{lum}}$ parameters: a greater $k_{\rm{lum}}$ value corresponds to higher rotational support within the galaxy.}}
    \label{fig:klum_vs_BT}
\end{figure}
As demonstrated by such examples naively interpreting \textsc{bang}'s visible components (i.e. bulge, inner disc, and outer disc) as the photometrically identified bulges and discs could lead to improper conclusions. Fig.~\ref{fig:klum_vs_BT}, which shows a comparison between \textsc{bang} decomposition and a purely photometric approach \citep[see e.g. PyMorph][for a photometric decomposition of MaNGA galaxies]{Sanchez2022}{}{}, indeed demonstrate an anti-correlation between photometrically identified bulges and the $k_{\rm{lum}}$ parameter. Galaxies with higher photometric $\rm{B/T}$ ratio are, on average, more dispersion supported (i.e. lower $k_{\rm{lum}}$) although with an increasing scatter for larger $\rm{B/T}$ ratios. Indeed, even at photometric B/T as high as 0.8 galaxies cover a wide range of kinematic properties (i.e. $0.15\lesssim k_{\rm{lum}} \lesssim 0.70$) with a number of objects still demonstrating strong rotational support (i.e. $k_{\rm{lum}} \gtrsim 0.70 $). Also in cases of dispersion-support ($k_{\rm{lum}}\lesssim0.15$) galaxies span a range of photometric B/T ratios with some objects retaining even more than half of their light in a photometrically identified disc component despite the absence of net rotation. This ultimately demonstrates some of the difficulties that could arise when interpreting photometric bulges and discs. A characterization of these components without the inclusion of the kinematics can be non-exhaustive since galaxies with the same value of $\rm{B/T}$ can reflect a variety of kinematic behaviours. In these regards, our approach helps in a better characterization of galaxies structure and kinematics although, it is still not clear \fr{how} far this can be pushed. For instance, it is still to be understood if \textsc{bang} can fully identify and characterise pseudobulges in some well-defined region of its parameter space. In the future, we are planning to apply our approach to simulations where decomposition can be performed by exploiting the full 6D phase space information of the stars \citep[][]{Du2020,Zana_2022}{}{}.

\begin{figure*}
    \centering
    \includegraphics[width=1.05\textwidth,height=0.75\textheight,trim={2cm 0cm 0 0}]{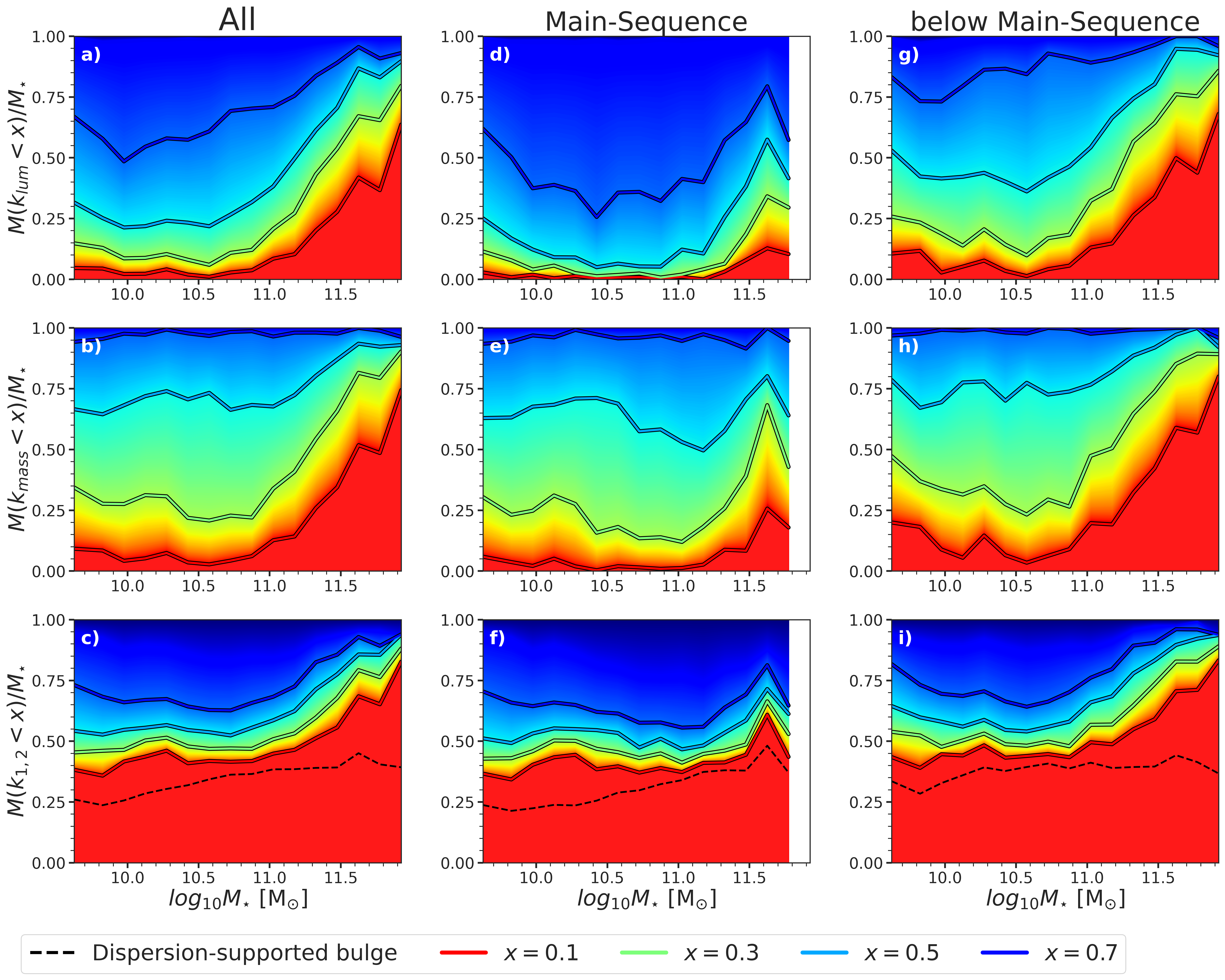}
    \caption{The horizontal axes show bins of stellar mass while the vertical axes are the mass fraction as a function of different kinematic tracers (i.e. from top to bottom: $k_{\rm lum}$, $k_{\rm mass}$ and $k_1$, $k_2$). The colour-coding from red to blue indicates an increase in the rotational support. The left column refers to the whole sample, the middle column includes only "below main-sequence" galaxies (i.e. $\Delta \rm MS <-0.5~\rm dex$) while the right column considers only main-sequence systems (i.e. $\Delta \rm MS >-0.5~\rm dex$). The bottom legend shows the different thresholds $x$ explicitly shown in the plot.}
    \label{fig:mass_budget}
\end{figure*}  

\section{Mass budget and its dependence on kinematics and star-formation}
\label{sec:mass-budget}

In this section, we aim to quantify the stellar mass budget of galaxies as a function of their kinematic state and distance from the star-forming main sequence (SFMS). Through this paper, we refer to as "main sequence" galaxies those systems above $-$0.5 ~\rm dex from the locus of the star-forming main sequence as defined in \cite{Fraser-McKelvie_2022}. Galaxies with lower SFR will be labelled as "below main sequence" systems.

Before diving into the results, based on what was discussed in Sec.~\ref{sec:Global and Structural kinematics}, we highlight some points that must be accounted for in the forthcoming analysis. First, if the goal is to estimate the contribution of different kinematic components to the mass budget of galaxies, we can not just assume that $k_1$ and $k_2$ always trace different kinematic components. Indeed, in some cases, they may trace the same component and it is actually their value (and relative difference) that provides important hints on the internal kinematic structure of galaxies. Secondly, treating $k_1$ and $k_2$ separately or combining them into $k_{\rm lum}$ (or $k_{\rm mass}$) may provide different results, as $k_{\rm lum}$ and $k_{\rm mass}$ generally trace just the dominant kinematic components.  This is why, in the present section, we compare both methods, looking at the mass budget of galaxies by either using $k_{\rm lum}$ and $k_{\rm mass}$ or treating each component separately.  

The main results of our analysis are summarized in Fig.~\ref{fig:mass_budget}. Here we show the cumulative distribution of different kinematic components per bin of galaxy stellar mass. The first column shows the results for our full sample, whereas the middle and right columns focus on the main sequence or `below' main sequence samples, respectively. 

\begin{table*}
	\centering
	\caption{Summary table of the mass fractions for the different kinematic tracers (i.e. $k_{1,2}$, $k_{\rm mass}$ and $k_{\rm lum}$) for the whole sample, "below main-sequence" galaxies (i.e. $\Delta \rm MS<-0.5\rm dex$) and "main-sequence" galaxies (i.e. $\Delta \rm MS>-0.5\rm dex$).}
	\label{tab:mass_budget}
	\small{\begin{tabular}{cccccc} 
 		\hline\vspace{-0.75em}\\
        \multicolumn{6}{c}{All}\vspace{0.2em}\\
		\hline\vspace{-0.75em}\\
		$\log_{10}{M_{\star}/M_{\sun}}$ & $k_{\rm 1,2}$ ; $k_{\rm mass}$ ; $k_{\rm lum}=0$& $k_{\rm 1,2}$ ; $k_{\rm mass}$ ; $k_{\rm lum}<0.1$ & $k_{\rm 1,2}$ ; $k_{\rm mass}$ ; $k_{\rm lum}<0.3$ & $k_{\rm 1,2}$ ; $k_{\rm mass}$ ; $k_{\rm lum}<0.5$ & $k_{\rm 1,2}$ ; $k_{\rm mass}$ ; $k_{\rm lum}<0.7$ \vspace{0.2em}\\
        \hline\vspace{-0.75em}\\
        $[9.75-  10.25]$ & 0.27 ; - ; - & 0.42 ; 0.06 ; 0.03 & 0.48 ; 0.30 ; 0.10 & 0.55 ; 0.69 ; 0.24 & 0.67 ; 0.96 ; 0.55\\
        $[10.25-  10.75]$ & 0.34 ; - ; - & 0.42 ; 0.04 ; 0.02 & 0.48 ; 0.23 ; 0.08 & 0.54 ; 0.71 ; 0.23 & 0.64 ; 0.97 ; 0.61\\
        $[10.75-11.25]$ & 0.38 ; - ; - & 0.44 ; 0.10 ; 0.07 & 0.50 ; 0.30 ; 0.19 & 0.58 ; 0.69 ; 0.38 & 0.68 ; 0.98 ; 0.72\\
        $[11.25-11.75]$ & 0.40 ; - ; - & 0.56 ; 0.34 ; 0.27 & 0.66 ; 0.64 ; 0.51 & 0.77 ; 0.85 ; 0.70 & 0.86 ; 0.98 ; 0.88\vspace{0.2em}\\
        \hline\vspace{-0.75em}\\
        $[9.75- 11.75]$ & 0.35 ; - ; - & 0.45 ; 0.11 ; 0.08 & 0.52 ; 0.33 ; 0.20 & 0.59 ; 0.72 ; 0.36 & 0.69 ; 0.98 ; 0.69\vspace{0.2em}\\
		\hline\vspace{-0.85em}\\
 		\hline\vspace{-0.75em}\\
        \multicolumn{6}{c}{Below Main-Sequence}\vspace{0.2em}\\
		\hline\vspace{-0.75em}\\
		$\log_{10}{M_{\star}/M_{\sun}}$ & $k_{\rm 1,2}$ ; $k_{\rm mass}$ ; $k_{\rm lum}=0$& $k_{\rm 1,2}$ ; $k_{\rm mass}$ ; $k_{\rm lum}<0.1$ & $k_{\rm 1,2}$ ; $k_{\rm mass}$ ; $k_{\rm lum}<0.3$ & $k_{\rm 1,2}$ ; $k_{\rm mass}$ ; $k_{\rm lum}<0.5$ & $k_{\rm 1,2}$ ; $k_{\rm mass}$ ; $k_{\rm lum}<0.7$ \vspace{0.2em}\\
        \hline\vspace{-0.75em}\\
        $[9.75-  10.25]$ & 0.35 ; - ; - & 0.43 ; 0.10 ; 0.07 & 0.50 ; 0.33 ; 0.18 & 0.57 ; 0.72 ; 0.44 & 0.69 ; 0.99 ; 0.78\\
        $[10.25-  10.75]$ & 0.39 ; - ; - & 0.45 ; 0.06 ; 0.03 & 0.49 ; 0.28 ; 0.14 & 0.55 ; 0.75 ; 0.39 & 0.66 ; 0.99 ; 0.87\\
        $[10.75-11.25]$ & 0.40 ; - ; - & 0.47 ; 0.15 ; 0.10 & 0.53 ; 0.39 ; 0.28 & 0.63 ; 0.77 ; 0.58 & 0.74 ; 0.99 ; 0.91\\
        $[11.25-11.75]$ & 0.41 ; - ; - & 0.60 ; 0.41 ; 0.34 & 0.72 ; 0.72 ; 0.63 & 0.82 ; 0.92 ; 0.81 & 0.91 ; 0.99 ; 0.96\vspace{0.2em}\\
        \hline\vspace{-0.75em}\\
        $[9.75- 11.75]$ & 0.39 ; - ; - & 0.49 ; 0.17 ; 0.13 & 0.56 ; 0.42 ; 0.30 & 0.64 ; 0.79 ; 0.54 & 0.75 ; 0.99 ; 0.90\vspace{0.2em}\\
		\hline\vspace{-0.85em}\\
 		\hline\vspace{-0.75em}\\
        \multicolumn{6}{c}{Main-Sequence}\vspace{0.2em}\\
		\hline\vspace{-0.75em}\\
		$\log_{10}{M_{\star}/M_{\sun}}$ & $k_{\rm 1,2}$ ; $k_{\rm mass}$ ; $k_{\rm lum}=0$& $k_{\rm 1,2}$ ; $k_{\rm mass}$ ; $k_{\rm lum}<0.1$ & $k_{\rm 1,2}$ ; $k_{\rm mass}$ ; $k_{\rm lum}<0.3$ & $k_{\rm 1,2}$ ; $k_{\rm mass}$ ; $k_{\rm lum}<0.5$ & $k_{\rm 1,2}$ ; $k_{\rm mass}$ ; $k_{\rm lum}<0.7$ \vspace{0.2em}\\
        \hline\vspace{-0.75em}\\
        $[9.75-  10.25]$ & 0.23 ; - ; - & 0.41 ; 0.04 ; 0.01 & 0.48 ; 0.28 ; 0.06 & 0.54 ; 0.68 ; 0.13 & 0.66 ; 0.96 ; 0.42\\
        $[10.25-  10.75]$ & 0.27 ; - ; - & 0.39 ; 0.01 ; 0.01 & 0.46 ; 0.17 ; 0.02 & 0.52 ; 0.67 ; 0.06 & 0.61 ; 0.97 ; 0.32\\
        $[10.75-11.25]$ & 0.34 ; - ; - & 0.39 ; 0.01 ; 0.01 & 0.44 ; 0.15 ; 0.02 & 0.49 ; 0.55 ; 0.08 & 0.57 ; 0.96 ; 0.37\\
        $[11.25-11.75]$ & 0.40 ; - ; - & 0.46 ; 0.12 ; 0.07 & 0.50 ; 0.37 ; 0.15 & 0.58 ; 0.65 ; 0.35 & 0.68 ; 0.94 ; 0.63\vspace{0.2em}\\
        \hline\vspace{-0.75em}\\
        $[9.75- 11.75]$ & 0.30 ; - ; - & 0.40 ; 0.03 ; 0.01 & 0.46 ; 0.20 ; 0.04 & 0.52 ; 0.63 ; 0.11 & 0.61 ; 0.96 ; 0.39\vspace{0.2em}\\
		\hline
	\end{tabular}}
\end{table*}

Starting from  Panel a),  we show 
the cumulative distribution of mass as a function of $k_{\rm lum}$, with highlighted arbitrary thresholds of $0.1,0.3,0.5,0.7$ to guide the eye. The colour coding maps the threshold levels on a continuous scale and it visually helps in the interpretation and comparison of the different panels\footnote{We arbitrarily choose red for high dispersion support (i.e., $x=0.1$) and dark blue for rotational support (i.e., $x=1.0$).}. The diagram is built as follows. First, for each galaxy falling in one of the stellar mass bins we compute $k_{\rm lum}$ following Eq.~\ref{eq:k_lum}. Then, given a threshold $x$, the mass fraction within $x$ is computed by summing the stellar mass of all those galaxies with $k_{\rm lum}$ below $x$ and dividing by the total stellar mass in the bin; we refer to Tab.~\ref{tab:mass_budget} for the exact values of the mass fraction\footnote{For reasons of clarity we report the fractions in larger mass bins as compared to Fig.~\ref{fig:mass_budget}. The values with the finer binning are available on request}. Note that this is a cumulative way of computing the mass fraction which, by construction, adds up to 1 when $x=1.0$. Essentially, for a fixed stellar mass, given that $k_{\rm lum}$ traces the total rotational support of the galaxy, a higher presence of "cold" colours in the vertical direction means that more stellar mass lies in rotationally supported systems. Focusing on the red line (i.e. $k_{\rm lum}<0.1$), which refers to dispersion-supported systems, we note that it remains as small as few percents up to $M_{\star}\simeq10^{11.0}~M_{\sun}$, when it starts to rise, reaching up to $\simeq 60\%$ of the mass fraction for $M_{\star}\simeq10^{12.0}~M_{\sun}$. Even though the increase of the mass fraction in dispersion-supported systems with $M_{\star}$ is expected due to the presence of ellipticals, we highlight that panel a) clearly shows that even for high-mass galaxies a considerable ($\simeq30\%$) amount of mass resides in a, at least mildly, rotating component. This is quite in contrast with analysis relying on visual morphology, which tends to overestimate the importance of ellipticals and/or spheroids 
(i.e., $60\%$ of the stellar mass fraction already at $M_{\star}\simeq10^{11} M_{\sun}$ \citep[][]{Moffet_2016,Bellstedt_2023}{}{}). Sources of possible visual morphology misclassification in photometric studies due to the lack of kinematic data were already been pointed out by several authors \citep[][]{Krajnovi_2008,Cappellari_2011,Fraser-McKelvie_2022}{}{}; we provide in Appendix \ref{app:missclassified_ellipticals} two examples of visually classified elliptical galaxies with a clear rotational pattern. 


Below $M_{\star}\simeq10^{11}~M_{\sun}$ the mass budget is clearly dominated by dynamically cold discs, even though galaxies with intermediate rotation $(k_{\rm lum}<0.5)$ contribute up to $25\%$. The relevance of "intermediate" galaxies has already been reported by other authors \citep[e.g.][]{Fraser-McKelvie_2022}{}{}. In terms of the structure composing a galaxy, intermediate rotational support is supposed to be caused by the presence of thick discs, strong bars, prominent bulges, etc.\footnote{Note that, as it has recently been pointed out by \cite{Croom2021}, intermediate rotation can be also caused by disc fading. We refer to Sec.~\ref{sec:discussion} for a detailed discussion.} Unfortunately, our model does not account for such complexity and, right now, it is still impossible to disentangle between all of these possible cases. Particularly interesting are thick discs, whose presence is supposed to be ubiquitous in disc galaxies of any morphology. In our modelling, those are still described as razor-thin, an approximation which is less appropriate for highly inclined galaxies. Since in our sample, highly inclined galaxies are less than a few percent we do not expect such an effect to play a significant role. It is not clear whether thick discs can already be identified by our method, for instance, by looking at the $k_1$ and $k_2$ parameters. Despite that, given the promising results obtained so far, we plan to include those components in our modelling as we expect to be able to disentangle among them in future higher-resolution IFS surveys. Interestingly, the very low mass bins seem to point toward a slight increase in the dispersion support compared to intermediate masses. Dwarf galaxies are expected to be more turbulent systems and similar results have already been reported in the literature \citep[for instance with the CALIFA survey,][]{Zhu_2018b}{}{}. However, at such small masses low signal-to-noise ratios and a stellar velocity dispersion of the order of the instrumental resolution can play a role and further investigation is needed to assess whether this trend extends below $M_{\star}<10^{9.5}~M_{\sun}$. Overall, this panel provides a picture consistent with what recently obtained by IFS analysis focused on integrated values of stellar spin ($\lambda_R$) or $V/\sigma$, as $k_{\rm lum}$ ultimately provide the same type of information: an integrated, luminosity average estimate of the kinematic of galaxies \citep[][]{van_de_Sande_2017,Guo_2020,Fraser_2022}{}{}.

Panel b) is built following the same steps as panel a) with the only difference that we replaced $k_{\rm lum}$ with the mass-weighted estimator $k_{\rm mass}$. Similarly to panel a), there is a clear trend of increasing mass fraction in dispersion-dominated systems with increasing $M_{\star}$. Overall, the qualitative behaviour of the two plots is similar, even though the colour coding in panel b) is shifted toward redder colours, implying a more significant role of dispersion in the overall mass budget. This is expected since, in the majority of the cases, the $M/L$ ratio in the inner regions (i.e. bulge and inner disc), where the galaxy has a stronger dispersion support, is larger than in the outer disc. As a result $k_{\rm mass}$ is almost always smaller than $k_{\rm lum}$ by construction. This also implies that fully rotationally-supported systems are extremely rare when we consider mass-weighted estimates.

Panel c) is built in a slightly different way. Here, we compute the mass fraction by accounting for the different components of the model (i.e. dispersion-supported bulge, inner disc and outer disc) separately. More precisely, we start again by selecting galaxies according to different total stellar mass bins (i.e. horizontal axis of the panel). Then for every galaxy in the mass bin, we compute the mass-weighted cumulative mass distribution of k, where the weight is used by combining the mass of all components having a certain k value. 
Note that, since in our assumptions the bulge is fully dispersion-supported, its mass is automatically included for any threshold $x$ (see the black dashed line in the plot). For instance, if a galaxy's inner and outer disc are both dispersion-dominated (i.e. $k_1<0.1$ and $k_2<0.1$) then all the galaxy's stellar mass will contribute to the first mass fraction bin (i.e. red line of the panel); on the contrary, in a case where only the outer disc is rotationally-supported (i.e. $k_1<0.1$ and $k_2>0.7$) then only the classical bulge and the inner disc fall in the first mass fraction bin (i.e. $x<0.1$, red region of the plot) whereas the outer disc contributes to the last mass fraction bin (i.e. $x>0.7$, over the blue line of the panel).

Similarly to panel a) and panel b), we see an increase in the mass fraction of dispersion-dominated structures with galaxy stellar mass. Interestingly, the trend is mostly caused by an increase in the fraction of dispersion-dominated (i.e. $k_1<0.1$ and/or $k_2<0.1$) inner and outer discs while the dispersion-supported bulge component (dotted line) remains almost on a constant level. As we discussed before, the increase in the fraction of dispersion-dominated inner and outer discs with $M_{\star}$ is not unexpected, since it is mostly related to the presence of galaxies which are overall pressure-supported. Below $M_{\star}\simeq10^{11}~M_{\sun}$ the mass fraction in a dispersion inner component (i.e. $k_{1,2}<0.1$) remains nearly constant at a level of $\simeq 40\%$. Similarly, the mass in the dispersion-supported bulge does not show significant trends with stellar mass roughly contributing to $\simeq 30\%$ of the mass fraction. There are two key differences between panels c) and a) or b). First, the contribution of dispersion-supported structures is significantly higher, in particular below $M_{\star}<10^{11.5}M_{\sun}$. This is simply because dispersion-supported structures rarely dominate the mass budget of an individual galaxy. As such, their presence is often washed out by integrated estimates, although once combined they can easily contribute to $30\%-40\%$ of the total mass budget. Secondly, the fraction of mass in "intermediate" structures ($0.1<k_{1,2}<0.7$) is significantly lower ($\lesssim 20\%$), with most of the mass being in either dispersion-supported or rotation-supported systems. Again, this is because averaging a pure bulge and a pure disc artificially creates a system with intermediate $k$. It is important to stress that this is not driven by our technique, as we actually allow for $k$ to cover every possible parameter. Instead, it nicely shows that our approach quantitatively confirms the generally accepted picture that galaxies are composed of two dominant and distinct structures: one almost entirely rotation-supported and the other dispersion-supported.

\begin{figure*}
    \centering
    \includegraphics[width=1.0\textwidth,height=0.5\textheight,trim={0 0.cm 0 0}]{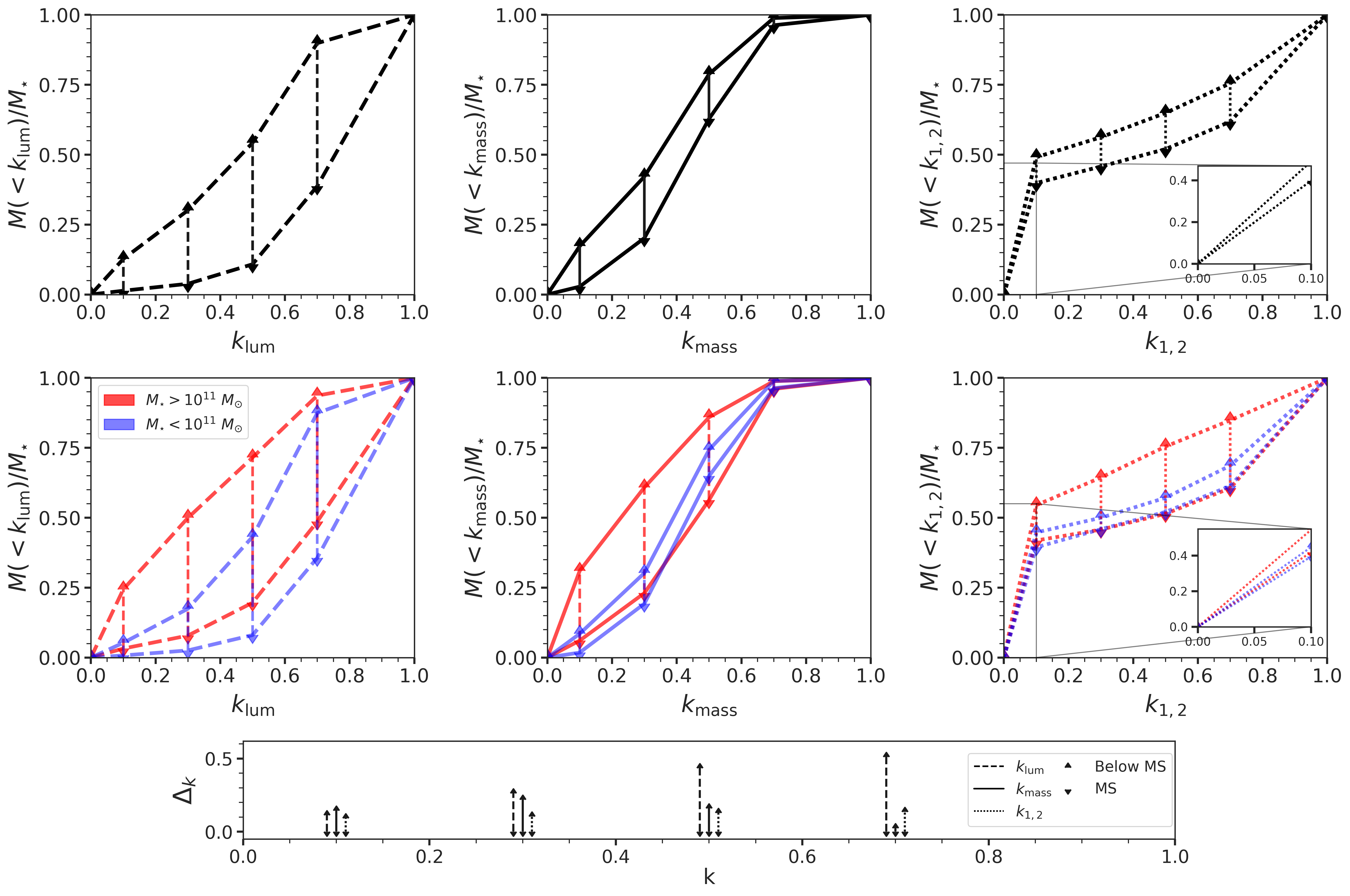}
    \caption{Cumulative mass fraction of $k_{\rm lum}$ (left, dashed line), $k_{\rm mass}$ (middle, solid line) and $k_{1,2}$ (right, dotted line). Upwards triangles show "below main-sequence" galaxies while the downwards triangles refer to "main-sequence" galaxies. The three panels on the top row refer to the whole sample while the middle row differentiates between galaxies with $M_{\star}>10^{11}M_{\sun}$ in red and galaxies with $M_{\star}<10^{11}M_{\sun}$ in blue. The panel in the bottom row illustrates the difference between "main-sequence" and "below main-sequence" galaxies (i.e. $\Delta_k$) for the whole sample as a function of $k$.}
    \label{fig:resume_plot}
\end{figure*}

We now proceed with the discussion by analysing and comparing the middle and the right columns of Fig.~\ref{fig:mass_budget}. We build these plots following, row by row, the same procedure described before dividing the sample into "main-sequence" galaxies (i.e. $\Delta\rm MS>-0.5\rm ~ dex$) and "below main-sequence" (i.e. $\Delta\rm MS<-0.5\rm ~ dex$) galaxies.

As expected, the dispersion-supported mass fraction at high masses is dominated by "below main-sequence" galaxies. Indeed, panels g)-h)-i) show a significant increase in the region below $x<0.1$, where $x$ can either be $k_{\rm lum}$, $k_{\rm mass}$ or $k_{1,2}$, of the plot with increasing mass, especially above $M_{\star}\simeq10^{11}M_{\sun}$, except for a possible tiny spike at $M_{\star} > 10^{11.5}\rm M_{\sun}$. Such spike could be due to poor statistics: indeed, the last mass bin is empty since there were no star-forming galaxies at such high masses in our sample\footnote{We considered mass bins with at least 10 galaxies in them.}. If we focus below $M_{\star}\simeq10^{11}M_{\sun}$ we see a striking difference between panel d) and panel g): 
the majority of galaxy mass at high $k_{\rm lum}$ is on the star-forming main-sequence. This is in line with what was found in many previous works \citep[]{Bell2012,Cappellari2013bulge_env,Omand2014,Lang2014,Kawinwanichakij2017}, and it has been used as evidence for the need to invoke structural transformation during or after quenching. However, this picture is challenged if we focus on panels e) and h) where the difference is much less evident when using $k_{\rm mass}$ as a tracer. In this case, for example, the mass fraction with $k_{\rm mass}<0.5$ is very similar for "main-sequence" and "below main-sequence" galaxies. This behaviour may be hinting at an evolution away from the SFMS which does not involve significant modification of the galaxy structure. More explicitly, since most of the star formation happens in the galactic disc, the process of quenching mostly affects its luminosity, and therefore its $M/L$ ratio. Moreover, in cases when the quenching is not caused by strong gravitational interactions, there are no particular reasons why it should imply a strong change in the dynamical structure of the galaxy and this is most likely why we do not see significant differences below $M_{\star}\simeq10^{10}M_{\sun}$ comparing panel e) and panel h). This result is supported, and reinforced, by panels f) and i), which may provide some insights into the structural transformation of galaxies. Above $M_{\star}\simeq10^{11}~M_{\sun}$, there is clearly an increasing importance of dispersion-supported structures, suggesting that quenched galaxies have experienced significant gravitational perturbation. Below this mass, the difference in mass stored in rotationally-supported and dispersion-supported structures is minimal, of the order of $10\%$ or less. This suggests that passive galaxies have undergone a limited structural transformation during or after their quenching phase.

\section{Discussion}
\label{sec:discussion}


Fig.~\ref{fig:resume_plot} shows the cumulative stellar mass fraction as a function of $k_{\rm lum}$ (left panel, dashed lines), $k_{\rm mass}$ (middle panel, bold lines) and $k_{1,2}$ (right panel, dotted lines) for "main-sequence" (downwards triangle) and "below main-sequence" (upwards triangle) galaxies. The three panels on top refer to the whole sample while the middle row differentiates among galaxies above $M_{\star}\simeq10^{11}M_{\sun}$ in red colours and galaxies below $M_{\star}\simeq10^{11}M_{\sun}$ in blue colours. The bottom panel directly compares the difference (i.e. $\Delta_k$) between main-sequence and below main-sequence galaxies for the whole sample among different kinematic tracers. We reported as well in Tab.~\ref{tab:mass_budget_overall} the cumulative mass fractions for the whole sample and for galaxies below and above $10^{11}~M_{\sun}$ to quantitatively compare among main-sequence and below main-sequence galaxies\footnote{Note that all the fractions reported in Tab.~\ref{tab:mass_budget_overall} are directly
comparable to each other since to compute them we used the total stellar mass of the whole sample as a common denominator.}.

Focusing on $k_{\rm lum}$ first, we see a strong dependence on whether the galaxy is on the main sequence or below it, with "below main-sequence" galaxies more dispersion-supported than "main-sequence" galaxies. Interestingly, such difference becomes stronger for higher degrees of rotational support reaching a maximum of $\Delta_k \simeq 0.5$. When dividing according to stellar mass (bottom left panel of Fig.~\ref{fig:resume_plot}) we observe that massive galaxies are overall more dispersion-dominated than their less massive counterpart. $\Delta_k$ monotonically increases as a function of $k_{\rm lum}$ for galaxies below $M_{\star}\simeq 10^{11}M_{\sun}$ being as large as $\Delta_k \simeq 0.5$ at $k_{\rm lum}\simeq 0.7$.   

This trend is much reduced when employing mass-weighted estimators. While galaxies below the main sequence still demonstrate higher dispersion support than their main sequence counterparts, the difference is less significant and, opposite to what is seen for $k_{\rm lum}$, slightly diminishes at larger $k$-values. Similarly, when dividing according to stellar mass (bottom middle panel of Fig.~\ref{fig:resume_plot}) we observe an almost negligible $\Delta_k$ for galaxies below $M_{\star}\simeq 10^{11}M_{\sun}$, while the difference is larger for more massive galaxies. Interestingly, massive main-sequence galaxies show similar properties to their less massive counterparts.

\begin{table}
	\centering
	\caption{Cumulative mass fraction for the different kinematic tracers (i.e. $k_{1,2}$, $k_{\rm mass}$ and $k_{\rm lum}$) for "below main-sequence" galaxies (i.e. $\Delta \rm MS<-0.5\rm dex$) and "main-sequence" galaxies (i.e. $\Delta \rm MS>-0.5\rm dex$). We show three different mass bins, namely for the whole sample and above and below $M_\star=10^{11}M_\sun$. Note that all the fractions are directly comparable to each other since we used the total stellar mass of the whole sample as a common denominator. }
	\label{tab:mass_budget_overall}
	\small{\begin{tabular}{ccc} 
 		\hline\vspace{-0.75em}\\
        $k_{\rm 1,2}$ ; $k_{\rm mass}$ ; $k_{\rm lum}\leq x$ & Main-Sequence & Below Main-Sequence\vspace{0.2em}\\
        \hline\vspace{-0.75em}\\
        \multicolumn{3}{c}{$9.75\leq \log_{10}{M_\star/M_\sun}\leq 11.75$}\vspace{0.2em}\\
        \hline\vspace{-0.75em}\\
        $x=0.1$ & 0.17 ; 0.01 ; 0.01 & 0.28 ; 0.10 ; 0.07\\
        $x=0.3$ & 0.19 ; 0.08 ; 0.02 & 0.33 ; 0.25 ; 0.18\\
        $x=0.5$ & 0.22 ; 0.26 ; 0.04 & 0.38 ; 0.46 ; 0.32\\
        $x=0.7$ & 0.26 ; 0.40 ; 0.16 & 0.44 ; 0.58 ; 0.53\\
        $x=1.0$ & 0.41 ; 0.41 ; 0.41 & 0.59 ; 0.59 ; 0.59\vspace{0.2em}\\
		\hline\vspace{-0.85em}\\
 		\hline\vspace{-0.75em}\\
        \multicolumn{3}{c}{$9.75\leq \log_{10}{M_\star/M_\sun}\leq 11.00$}\vspace{0.2em}\\
		\hline\vspace{-0.75em}\\
        $x=0.1$ & 0.12 ; 0.01 ; 0.00 & 0.16 ; 0.03 ; 0.02\\
        $x=0.3$ & 0.14 ; 0.06 ; 0.01 & 0.18 ; 0.11 ; 0.06\\
        $x=0.5$ & 0.16 ; 0.20 ; 0.02 & 0.21 ; 0.27 ; 0.16\\
        $x=0.7$ & 0.19 ; 0.30 ; 0.11 & 0.25 ; 0.36 ; 0.31\\
        $x=1.0$ & 0.31 ; 0.31 ; 0.31 & 0.36 ; 0.36 ; 0.36\vspace{0.2em}\\
		\hline\vspace{-0.85em}\\
 		\hline\vspace{-0.75em}\\
        \multicolumn{3}{c}{$11.00\leq \log_{10}{M_\star/M_\sun}\leq 11.75$}\vspace{0.2em}\\
		\hline\vspace{-0.75em}\\
        $x=0.1$ & 0.04 ; 0.01 ; 0.00 & 0.12 ; 0.07 ; 0.06\\
        $x=0.3$ & 0.05 ; 0.02 ; 0.01 & 0.15 ; 0.14 ; 0.11\\
        $x=0.5$ & 0.05 ; 0.06 ; 0.02 & 0.17 ; 0.20 ; 0.16\\
        $x=0.7$ & 0.06 ; 0.09 ; 0.05 & 0.19 ; 0.23 ; 0.21\\
        $x=1.0$ & 0.10 ; 0.10 ; 0.10 & 0.23 ; 0.23 ; 0.23\vspace{0.2em}\\
		\hline
	\end{tabular}}
\end{table}

Looking at the right panels of Fig.~\ref{fig:resume_plot} we see that about $40\%$ of the mass fraction lies below $k_{1,2}\leq0.1$. This is due to the common presence of a dispersion-supported inner region in the galaxies of our sample. At intermediate values of $k_{1,2}$, we do not observe a significant increase in the mass fraction as most of the mass is in dispersion-supported (i.e. $k_{1,2}<0.1$) or rotationally-supported (i.e. $k_{1,2}>0.7$) components. This behaviour is particularly evident when focusing on galaxies below $M_{\star}\simeq10^{11}M_{\sun}$ (bottom right panel of Fig.~\ref{fig:resume_plot}). Note that, similarly to what observed for $k_{\rm mass}$, $\Delta_k$ is significantly smaller for less massive galaxies than their more massive counterpart. 

All of these findings have important implications for the interpretation of our results in a broader context. Luminosity-weighted kinematic tracers not only capture changes in a galaxy dynamics but also in its SFR. While this may be informative in certain scenarios, it can lead to biased interpretations of the relationship between kinematics and quenching. In this context, the decrease in rotational support should be viewed more as a consequence than a cause of quenching. As galaxies deviate from the SFMS, their discs cease star formation, leading to a decrease in luminosity (and an increase in the $M/L$ ratio). Consequently, given that the disc retains most of the galaxy's angular momentum, a galaxy below the main sequence may exhibit an apparent reduction in (luminosity-weighted) rotation support compared to another "main-sequence" galaxy with similar intrinsic kinematics. Indeed, most of the difference between luminosity- and mass- weighted results is for main-sequence galaxies. Considering that many studies on galaxy dynamics rely on light-weighted properties such as $V/\sigma$ or $\lambda_R$ \citep[][]{Cappellari_2007,Emsellem2007,Zhu2024}{}{}, caution is advised, as our findings indicate the difficulty of disentangling trends in galaxy kinematics from variations in the underlying stellar population when relying on light-weighted tracers. This caution extends to the assumption of a constant stellar $M/L$ ratio commonly made in dynamical models \citep[][]{Zhu_2018,Santucci2022,Jin2023}{}{}. Our modelling incorporates, through three distinct $M/L$ ratios (one for each visible component), a radial variation of the galaxy's average $M/L$ ratio enhancing the disparity in behaviours between $k_{\rm lum}$ and $k_{\rm mass}$.
It is important to note that assuming a constant $M/L$ ratio would have led to identical definitions for $k_{\rm lum}$ and $k_{\rm mass}$, effectively washing out any distinction between light-weighted and mass-weighted quantities. This highlights the need to consider radial gradients of the $M/L$ ratio in dynamics a fact that has recently been pointed out in \cite{Liang2023} and \cite{Chu2023}.

Focusing on the physical interpretation of our findings, we conclude that the small spread in the mass-weighted $M(<k)/M_*$ (compared to its luminosity-weighted counterpart, see middle and right panels of Fig.~\ref{fig:resume_plot}) and its decrease with increasing galaxy rotational support point towards a little structural evolution as galaxies depart from the SFMS. Below  $M_{\star}<10^{11}M_{\sun}$, a connection between galaxy quenching and structural/morphological transformation is less strong than what was previously thought. Instead, the predominant factor at play appears to be associated with disc fading. Such a scenario has already been advocated by several authors \citep{Blanton2005,Cortese2009,Woo2017,Cortese2019,Croom2021}. 

This can have important implications in the context of the evolution of S0 galaxies. Lenticular galaxies, in comparison to spirals, typically exhibit reduced rotational support. For instance, S0 galaxies follow a Fall relation \citep[i.e. stellar specific angular momentum vs stellar mass,][]{Fall1983}{}{} parallel to that of spirals but for a lower intercept and hence, a lower specific angular momentum at fixed mass. However, when these galaxies are decomposed into their bulge and disc components the picture changes. Discs of lenticular galaxies follow the same Fall relation as that of spirals \citep[][]{Rizzo2018}{}{} suggesting a structural similarity between the two. The lower rotational support of lenticulars is not a consequence of a strong modification of the galactic disc. Instead, it is likely a consequence of a change in the ratio between their inner dispersion-dominated component and their rotationally supported disc component. This interpretation gains further support from our analysis. When the disc of the progenitor of an S0 galaxy ceases star-formation its light dims and, as a consequence, its light-weighted kinematics decreases toward a lower rotational support. 

Interestingly studies examining the individual spectra of bulges and disc in S0s \citep[see e.g.][]{Fraser-McKelvie2018,Dominguez2020,Johnston2022}{}{} unveiled a dependence of the assembly history of these components on stellar mass. Bulges within galaxies with stellar masses exceeding $\gtrsim 10^{10}\rm M_{\sun}$ assemble most of their mass early in their lifetimes, whereas those in lower mass galaxies formed over more extended time-scales and more recently. Such observations have been interpreted as supporting evidence for the formation of massive S0 galaxies occurring via an inside-out scenario, while in lower mass S0s, the bulges and discs either formed together or through an outside-in scenario\footnote{It is important to note that these conclusions were drawn from studies with relatively limited statistical samples of few hundred galaxies.}. Differently from this work, those studies relied on a luminosity-based bulge+disc decomposition and did not consider the separate kinematics of bulges and discs in their analysis so it is not clear if a change in behaviours should be evident also from our analysis. Interestingly, Fig.~\ref{fig:mass_budget} hints toward a moderate increase in the mass within mildly rotating discs (i.e. $k_{1,2}\simeq 0.5$) below $\lesssim10^{10}~M_{\sun}$, although a focused and joint analysis of the kinematics and the stellar population is required before drawing any conclusion on a possible correlation of the two effects. 




Nonetheless, indications of a minimal structural evolution persist even when considering mass-weighted quantities. Such behaviour may be due to different reasons. Here we speculate about two possible explanations. Firstly, we note, by comparing panels f) and i) of Fig.~\ref{fig:mass_budget} and by looking at the right panel of Fig.~\ref{fig:resume_plot}, that below $M_{\star}<10^{11}~M_{\odot}$ the amount of mass in the dispersion-supported central region (i.e. $k_{1,2}<0.1$)  changes minimally for "main-sequence" and "below main-sequence" galaxies, while a larger difference can be seen at higher thresholds, namely $k_{1,2}=0.5$ and $k_{1,2}=0.7$.  From this observation, we infer that if structural modifications are indeed present, they likely predominantly influence the dynamics and structure of the galactic disc. This hypothesis gains support from simulations suggesting that the spheroidal mass component of galaxies is established early in the Universe, undergoing minimal changes after $z<2$ \citep[][]{Tacchella2019}{}{}. The increased dispersion support in "below main-sequence" galaxies may result from a thickening of the disc. This scenario is supported by the consideration that quenched galaxies, being generally older than their star-forming counterparts, may have attained a stable disc-like configuration in earlier epochs when discs were more turbulent, resulting in a thicker structure \citep[][]{Pinna2023}{}{}. Another possibility is that for fixed stellar mass, galaxies that grow their disc faster (and therefore assemble most of their mass at early times, being less star forming at $z=0$) are more likely to undergo bar instability~\citep[][]{Fraser_2020,Izquierdo2022}. Notably, stellar bars are dynamically stable structures that tend to increase the disc dispersion support~\citep[e.g.][and references therein]{2016MNRAS.459..199G}, contributing to the galactic morphological evolution \footnote{Note that bars are more commonly found in redder, less-star-forming galaxies suggesting a strong link between the presence of bars and galaxy quenching \citep[see, e.g.][]{Gavazzi2015}{}{} }.

Of course, these scenarios are not valid above $M_{\star}>10^{11}~M_{\sun}$, where our results point towards a different mechanism responsible for quenching massive galaxies. In this case, "below main-sequence" galaxies demonstrate a substantial change in their structure even though such process is segregated only at very high masses. The mass threshold of $10^{11}M_{\sun}$ is consistent with galaxies assembling more than $50\%$ of their stellar mass through mergers \citep{Oser_2010,Tacchella2019} and with the critical mass above which dry mergers relics dominate \citep[][]{Peng2010,Cappellari_2016}{}{}. Interestingly, we find at $M_\star\simeq10^{11.5}~M_\sun$ a mass fraction in dispersion-dominated structures (panel i) of Fig.~\ref{fig:mass_budget}) close to 0.7 consistent with the ex-situ mass fraction at $z=0$ reported in \cite{Tacchella2019}. This leads us to conclude that the importance of mergers in quenching galaxies and changing their morphology is confined to a very limited range of masses (i.e. $M_{\star}>10^{11}~M_{\sun}$).

\section{Summary and Conclusions}

In this work, we exploited the analysis presented in \citep[][]{Rigamonti_2023}{}{} with the publicly available code \textsc{bang} \citep[][]{Rigamonti_2022_software}{}{} on a morpho-kinematic decomposition of galaxies from the final data release of the MaNGA survey \citep[][]{Bundy_2015,Abdurro'uf_2022}{}. 

More specifically, we took advantage of \textsc{bang}'s ability to disentangle between the different galaxies components while characterising, through the parameters $k_1$ and $k_2$, their dynamical state (i.e. whether they are rotation- or dispersion- dominated). Moreover, by defining mass- or light- weighted kinematic tracers (i.e. $k_{\rm lum}$, $k_{\rm mass}$), we characterized the dynamical state of galaxies as a whole. Depending on these parameters ($k_{1}$, $k_2$, $k_{\rm mass}$ and, $k_{\rm lum}$) we were able to quantify in a generalized way the mass budget in the local Universe within the different galaxy components (i.e. dispersion-supported bulges and discs), for different type of galaxies (i.e. rotationally-supported or dispersion-supported) and depending on their location on the SFMS (i.e. "below main-sequence" galaxies: $\Delta\rm MS<-0.5\rm dex$ and "main-sequence" galaxies: $\Delta\rm MS>-0.5\rm dex$). The main results are reported in Tab.~\ref{tab:mass_budget_overall} and in Fig.~\ref{fig:mass_budget} and Fig.~\ref{fig:resume_plot} and can be summarised as follows:

\begin{itemize}
    \item In our sample, we found that $\simeq 8\%$ of the mass resides in dispersion-dominated galaxies (i.e. $k_{\rm lum}<0.1$\footnote{Note that this threshold is arbitrary. We reported in Tab.~\ref{tab:mass_budget_overall} the cumulative mass fractions for other thresholds.}). Most of this mass ($\simeq 6\%$) comes from massive (i.e. $M_\star>10^{11}~M_\sun$) below main-sequence galaxies. Galaxies with higher rotational support can still retain a significant fraction of their mass in a dispersion-dominated inner region which accounts for $\simeq 40-45\%$ of the mass in the whole sample.
    
    \item Above the critical mass of $M_{\star}\simeq10^{11}M_{\sun}$, the importance of passive, slowly rotating galaxies strongly increases. This result is independent of the technique used to differentiate between rotation- and dispersion- dominated systems. Following previous studies \citep[][]{Cappellari_2016,vandeSande2017,Tacchella2019}{}{}, we explain these trends in terms of mergers being the main mechanism shaping massive slow rotators and halting their star formation. Compared to previous results, based on photometric approaches \citep[][]{Gadotti_2009,Lange_2016,Moffet_2016,Bellstedt_2023}{}{}, our analysis segregates the role of mergers at higher masses highlighting the necessity of kinematic information to avoid misclassifications between spheroids and rotating passive systems \citep[][]{Emsellem2011,Krajnovic2013,Cortese2016}{}{}. 
    
    \item Below $M_{\star}\simeq10^{11}M_{\sun}$, galaxies span a wide, continuous range of $k_{\rm lum}$, $k_{\rm mass}$ or, similarly, of rotation-to-dispersion support. Interestingly, when looking at the galaxy different components, they naturally separate into a dispersion-dominated inner region and a rotationally-supported outer disc. Only a minor amount of mass resides in structures with intermediate rotation so that the overall galaxy dynamical support is determined by the relative weight of its dispersion-dominated and rotation-dominated components. This shows \textsc{bang}'s ability to decompose galaxies into physically motivated structures, stressing again the relevance of addressing and extending the bulge+disc problem from an IFS perspective.  
    
    \item Luminosity-based tracers of the kinematics encapsulate differences not only in the dynamical structure of a galaxy but also in their star formation history. Below $M_{\star}\simeq10^{11}M_{\sun}$, we observe a large difference in kinematic properties of galaxies as a function of their position in the star formation mass plane, when relying on luminosity-based tracers. The same difference is much less significant when considering mass-based quantities. Such difference between luminosity- or mass- weighted estimators is observable only thanks to our assumption of a spatially variable mass-to-light ratio. We conclude that relying on luminosity-based estimators can lead, in some circumstances, to a biased interpretation of the physical processes at play. Moreover, the assumption of a constant mass-to-light ratio commonly made in dynamical models can have a non-negligible impact on the estimated properties of galaxies.
    
    \item Given the small difference of the mass-weighted kinematics among "main-sequence" and "below main-sequence galaxies" we conclude that in the mass range $10^{9.75}M_{\sun}<M_{\star}<10^{11}M_{\sun}$, morphology and quenching have only a minor correlation as the main process at play is disc fading. Our work has indeed probed that, in the local Universe, there is no evidence for quenching being the cause or being caused by a morphological transformation.
\end{itemize}

\section*{Acknowledgements}
We thank the anonymous referee for their comments and suggestions that helped us to improve the quality of the paper.

MD acknowledge funding from MIUR under the grant
PRIN 2017-MB8AEZ, and financial support from ICSC – Centro
Nazionale di Ricerca in High Performance Computing, Big Data and
Quantum Computing, funded by European Union – NextGenera-
tionEU.

LC acknowledges support from the Australian Research Council Future Fellowship and Discovery Project funding scheme (FT180100066,DP210100337). Parts of this research were conducted by the Australian Research Council Centre of Excellence for All Sky Astrophysics in 3 Dimensions (ASTRO 3D), through project number CE170100013. 

FR acknowledges support via the ASTRO 3D science visitor program scheme and the CINECA award under the ISCRA initiative, for the availability of high-performance computing resources and support.

\section*{Data Availability}

The data underlying this article will be shared on reasonable request to the corresponding author.



\bibliographystyle{aa} 
\bibliography{main}




\appendix

\section{Missclassified Ellipticals }
\label{app:missclassified_ellipticals}

\begin{figure}
    \includegraphics[scale=0.225]{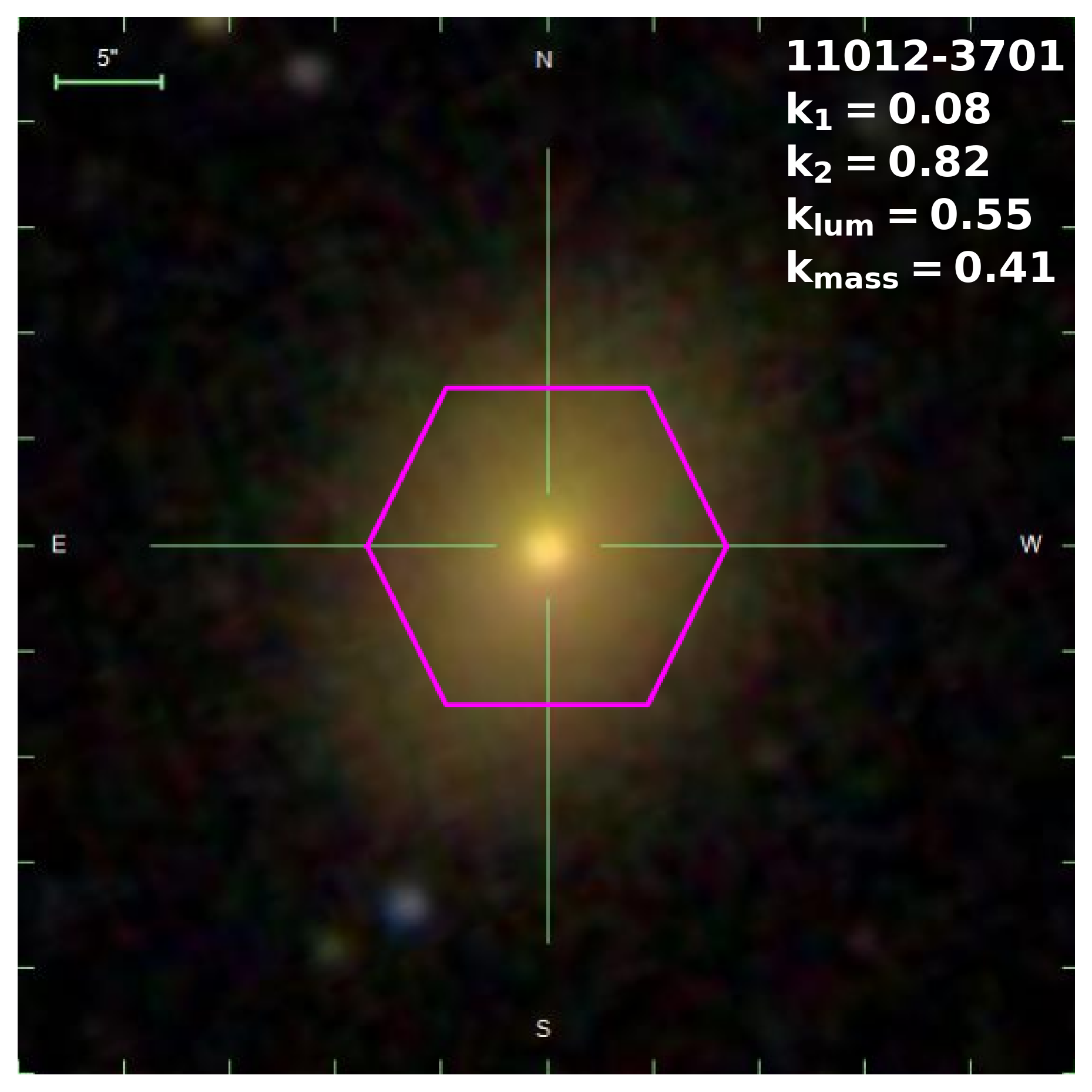}
    \includegraphics[scale=0.35,trim={0 1.3cm 0 0.0cm}]{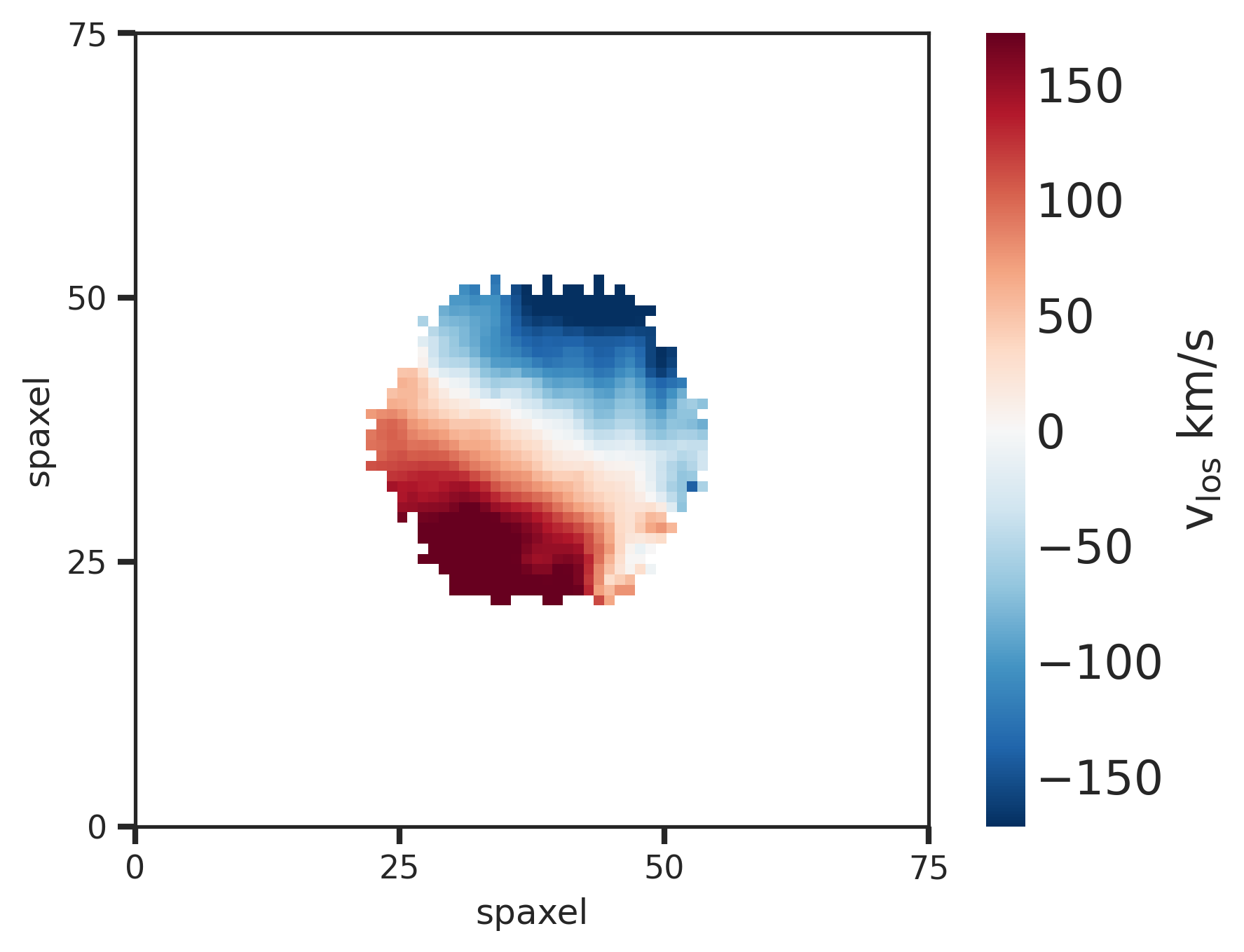}
    \vspace{0.02\textheight}
    
    \includegraphics[scale=0.225]{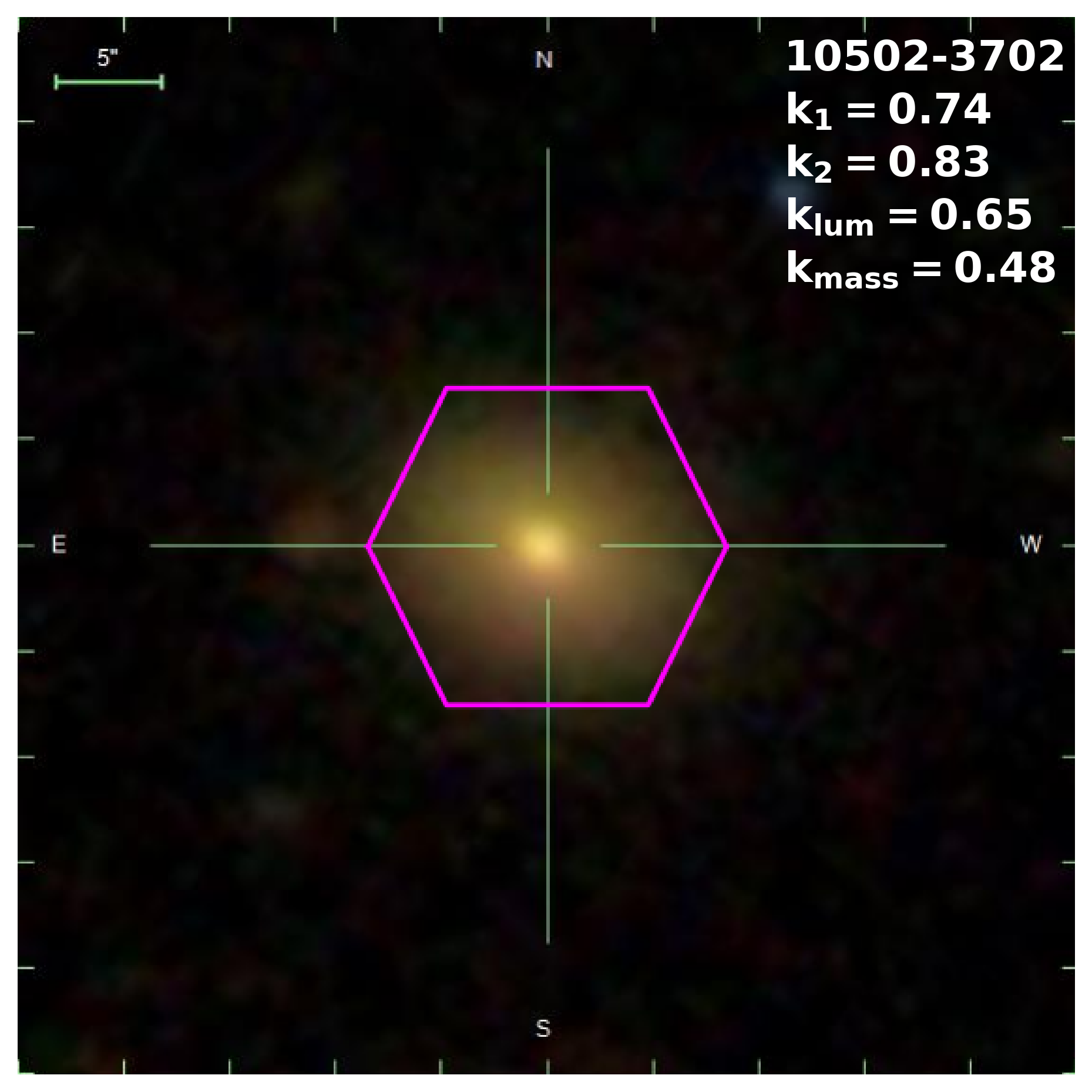}
    \includegraphics[scale=0.35,trim={0 1.3cm 0 0.0cm}]{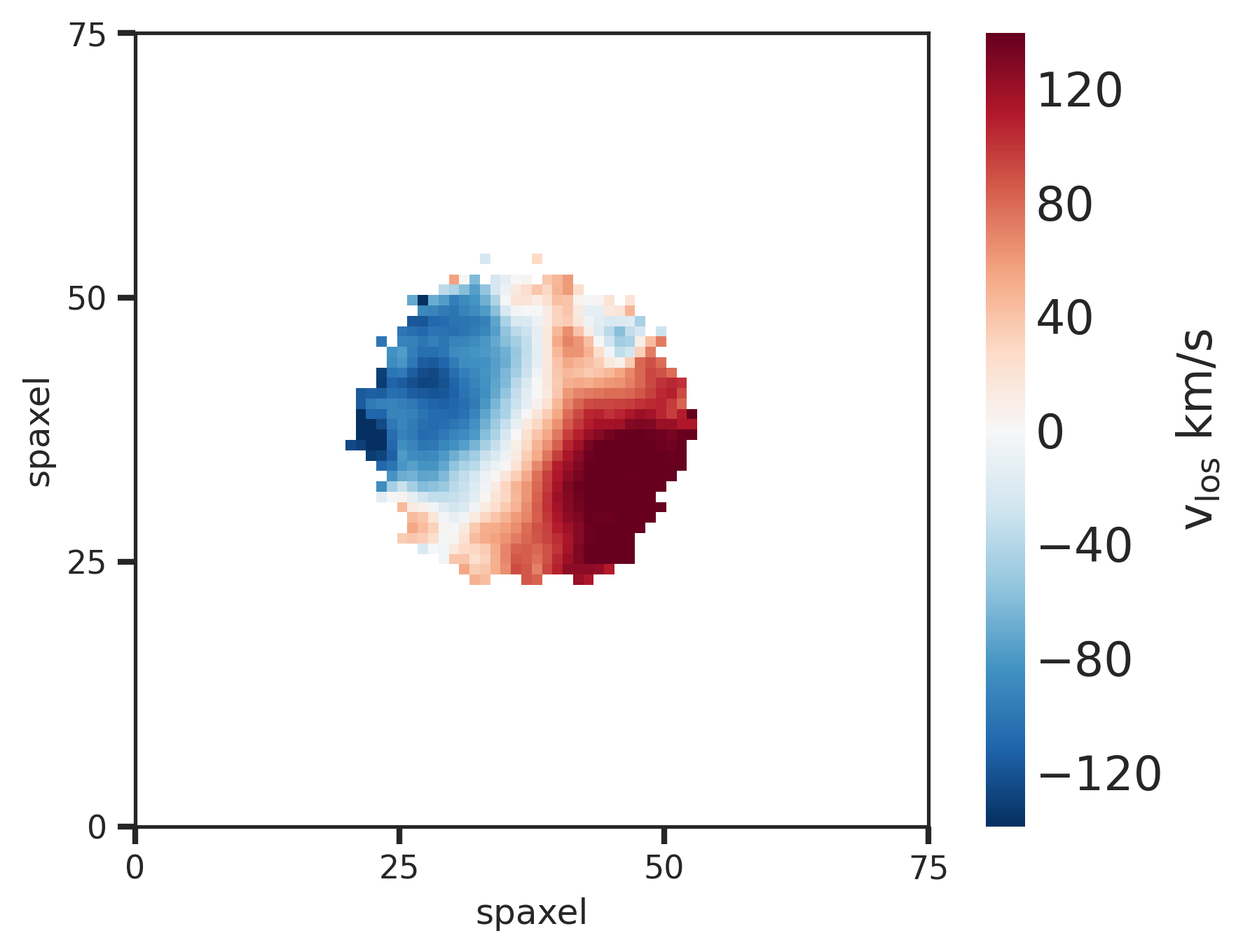}\vspace{0.02\textheight}
    
    \label{fig:missclassified_ellipticals}
    \caption{Two examples of galaxies (i.e. 11012-3701 top row, 10502-3702 bottom row) classified as ellipticals according to the VMC-VAC \citep[][]{Vazquez_2022}{}{} with a clear rotational pattern. }
    
\end{figure}

Fig.~\ref{fig:missclassified_ellipticals} shows two examples (11012-3701 top row, 10502-3702 bottom row) of galaxies classified as ellipticals according to the VMC-VAC \citep[][]{Vazquez_2022}{}{} that exhibit a clear rotational pattern.

11012-3701 has a stellar mass of $M_\star=10^{11.54}~M_\sun$ while 10502-3702 has a stellar mass of $M_\star=10^{10.83}~M_\sun$. In both cases $k_{\rm lum}$ indicates a quite strong rotational support.

\section{Comparison with other kinematic tracers}
\label{app:comparison_with_lambdaR}

In this section, we compare the kinematic tracer $k_{\rm lum}$ used in this work with the more commonly adopted $\lambda_R$. For the comparison we consider $\lambda_R$ estimates presented in \citep[][]{Fraser-McKelvie_2021}{}{} corrected for inclination and seeing.

The two parameters correlate with each other although some scatter is present. The correlation is close to linear, demonstrating the physical meaning of the $k_{\rm lum}$ parameter adopted throughout this work. We refer to Paper II for a comparison with $V/\sigma$.

\begin{figure}

\centering
\includegraphics[scale=0.5]{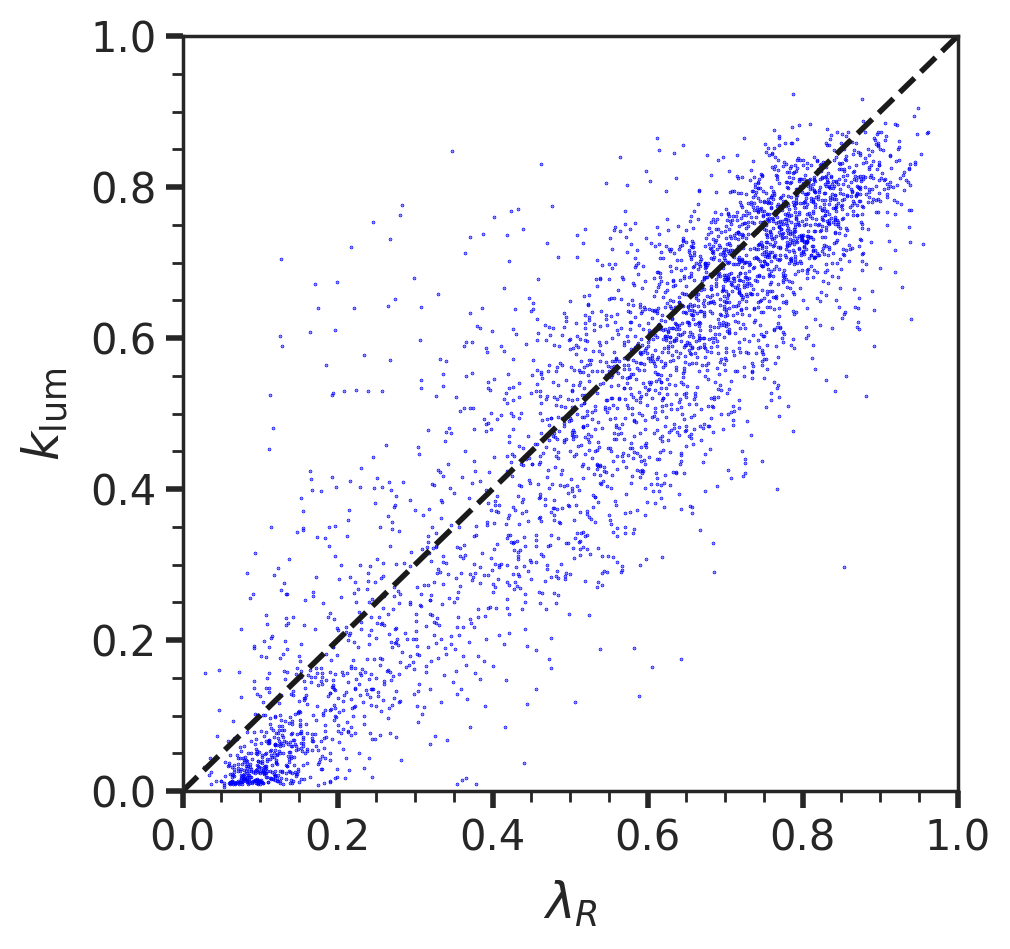}
\caption{Comparison between $k_{\rm lum}$ and $\lambda_R$. The blue dots are the data while the black dashed line is the one-to-one relation. For the comparison we consider $\lambda_R$ estimates presented in \citep[][]{Fraser-McKelvie_2021}{}{} corrected for inclination and seeing.}
\label{fig:klum_vs_lambdaR}

\end{figure}

\section{Further examples on k1 and k2 }
\label{app:other_k1_k2_examples}
Here we show two examples of galaxies with $k_1$ larger than $k_2$.

\begin{figure*}
    \includegraphics[scale=0.304]{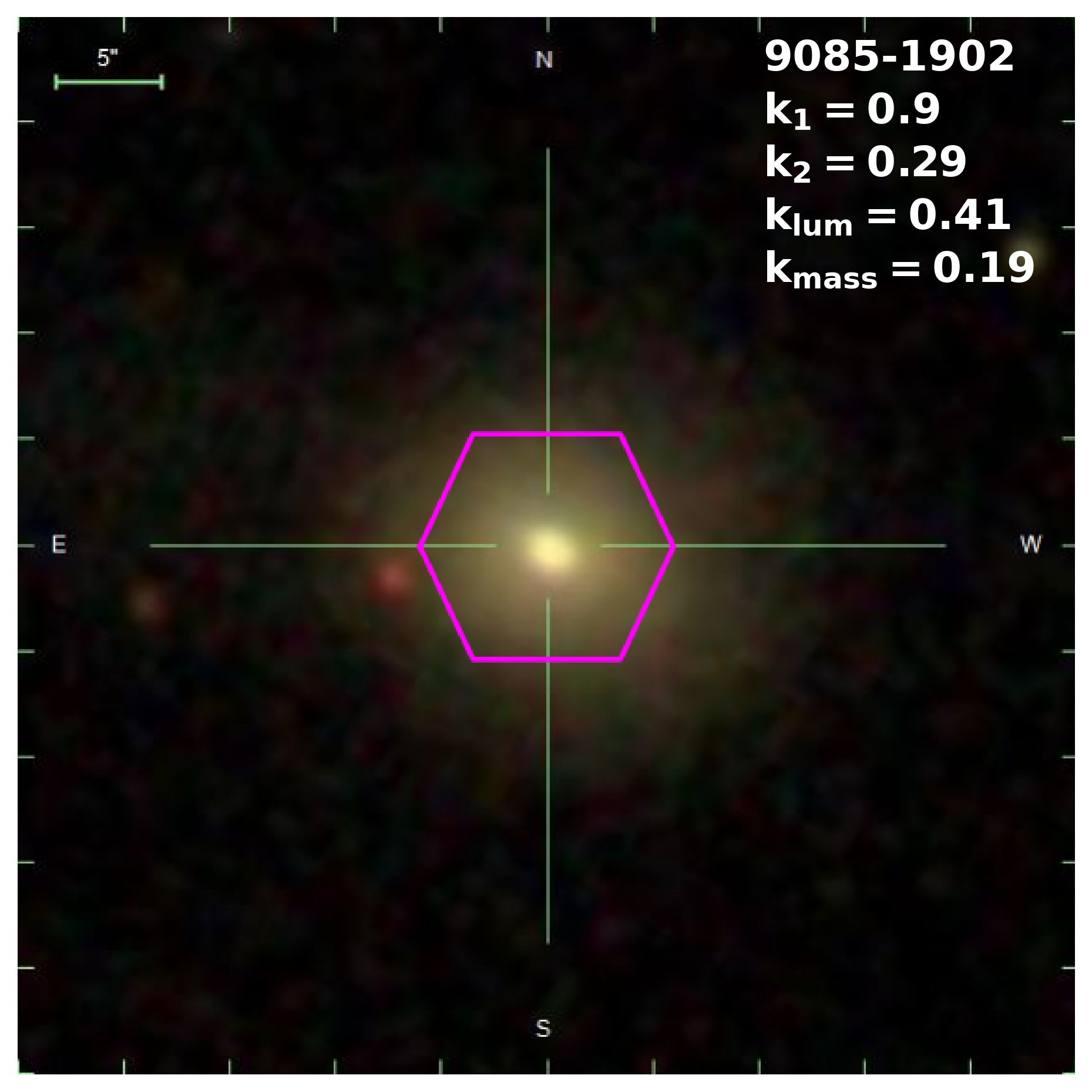}\hspace{0.015\textwidth}\includegraphics[scale=0.476,trim={0 1.3cm 0 0}]{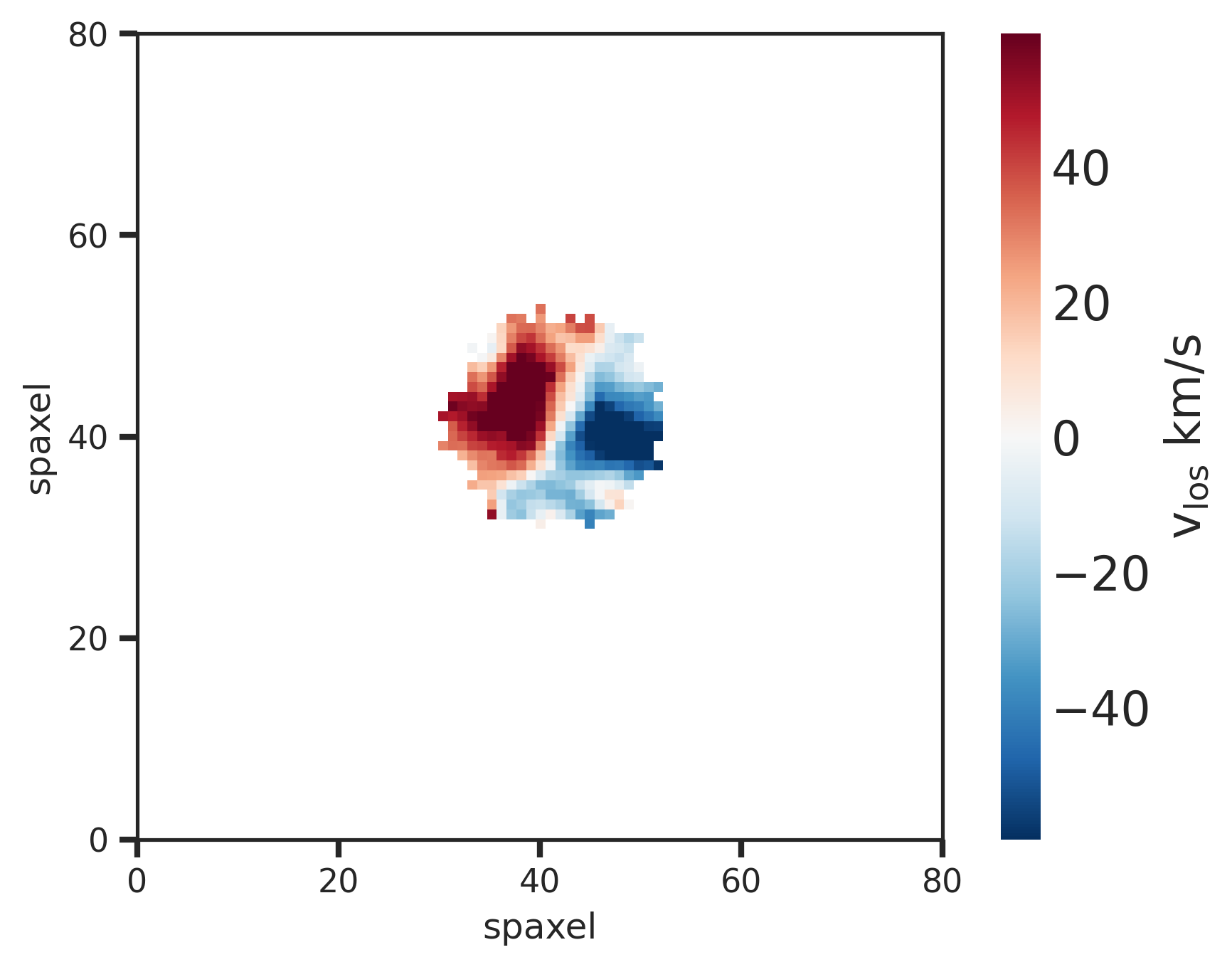}
    \includegraphics[scale=0.476,trim={0 1.3cm 0 0}]{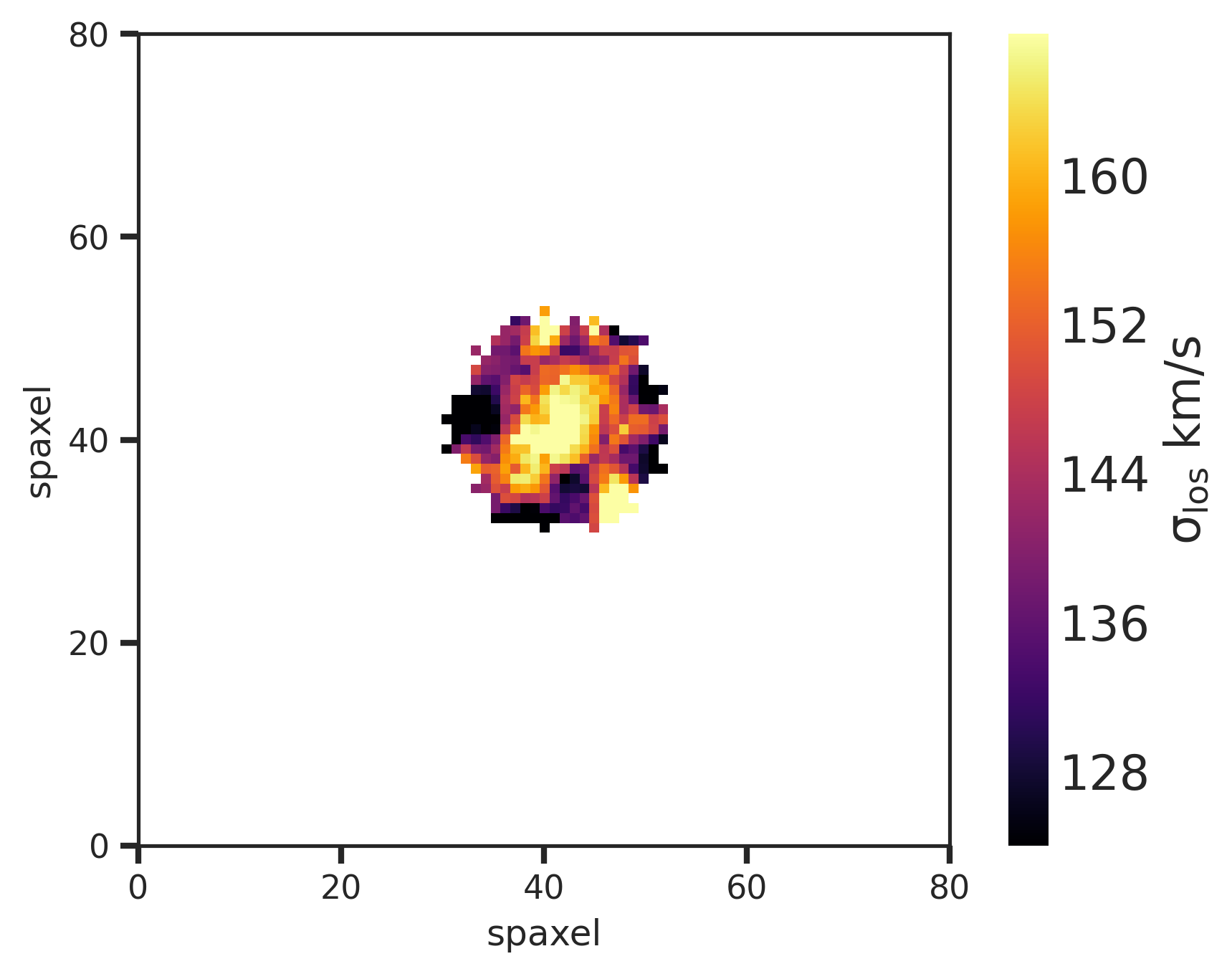}\vspace{0.02\textheight}\vspace{0.01\textheight}
    \includegraphics[scale=0.304]{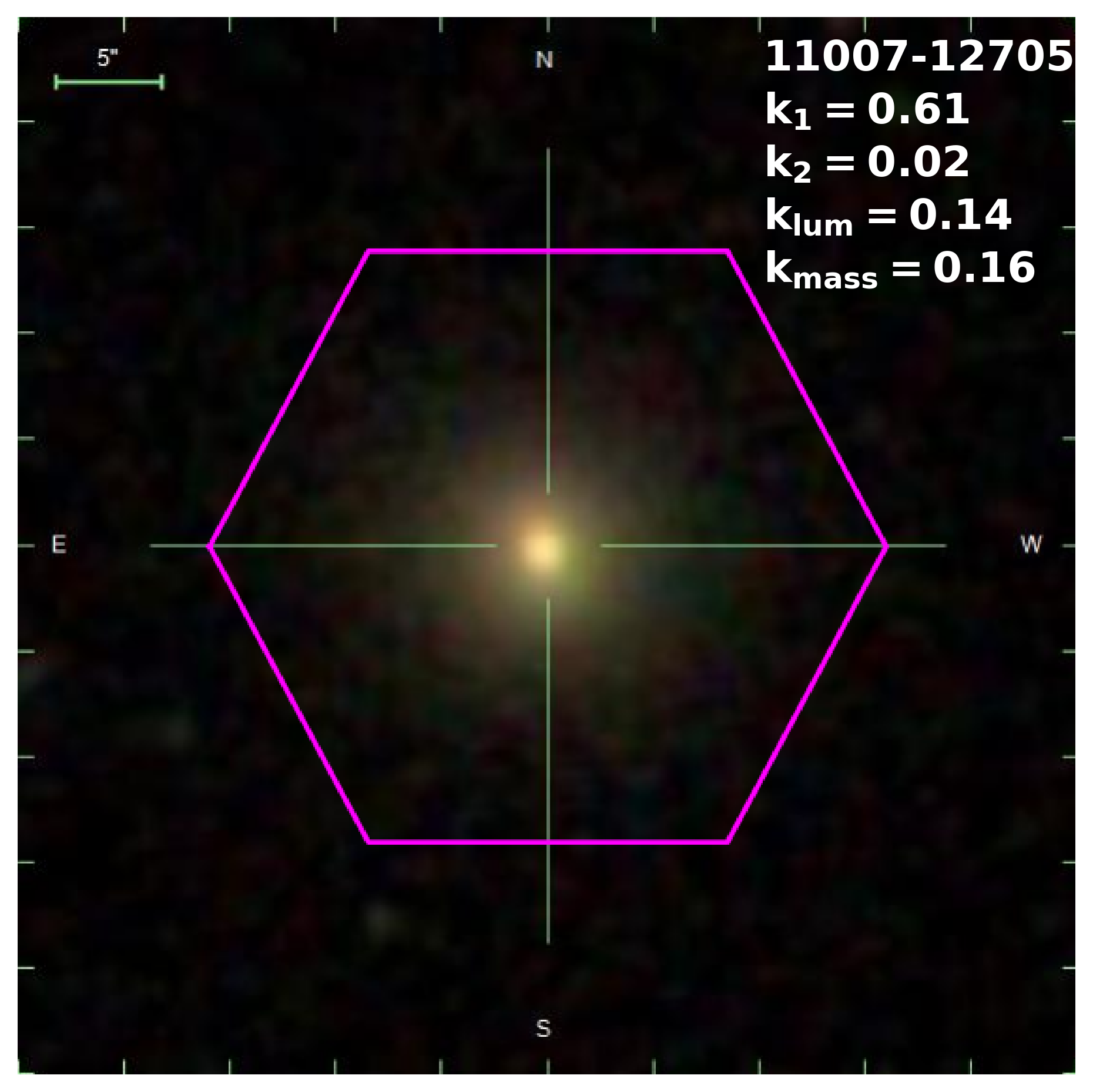}\hspace{0.015\textwidth}\includegraphics[scale=0.476,trim={0 1.3cm 0 0}]{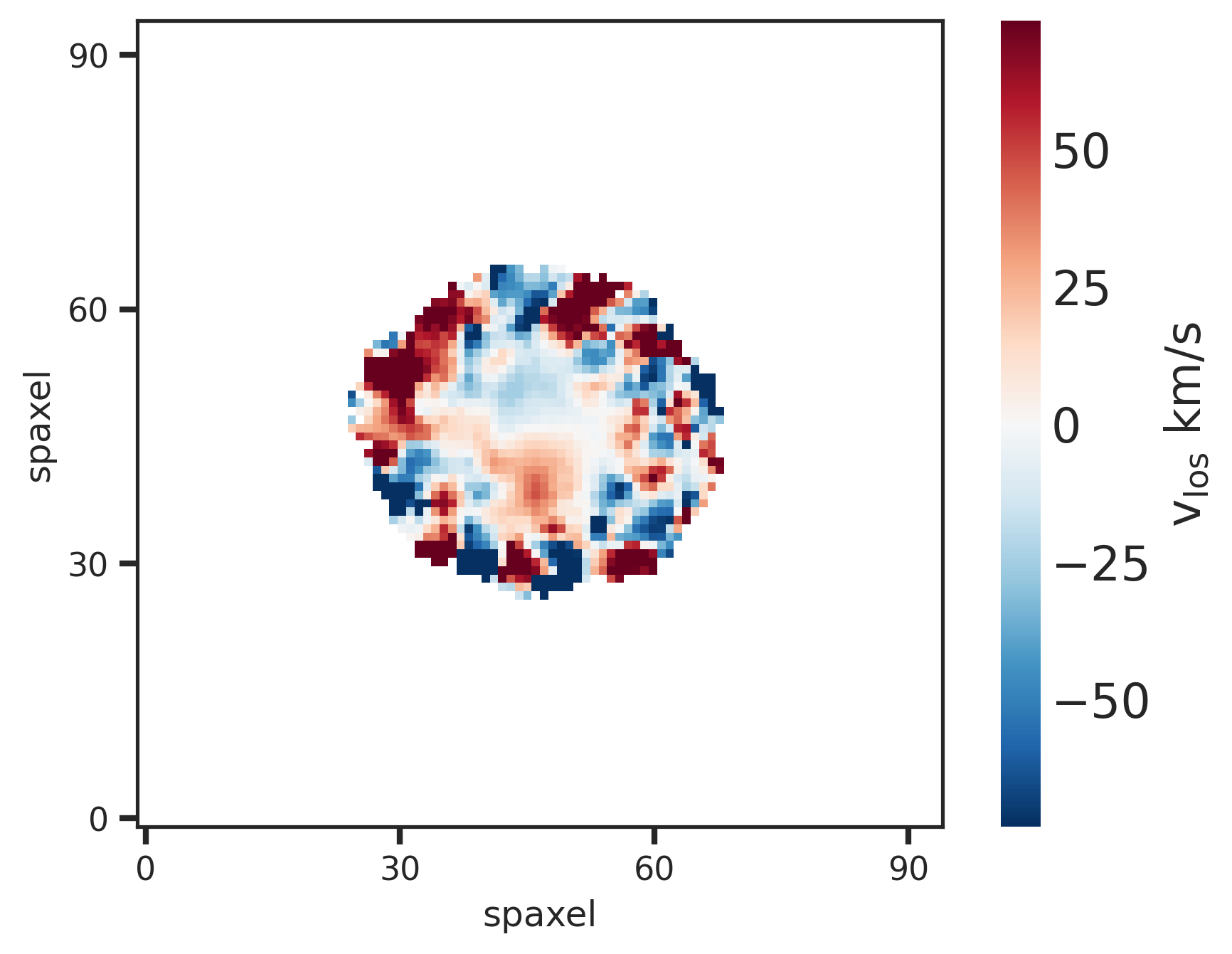}
    \includegraphics[scale=0.476,trim={0 1.3cm 0 0}]{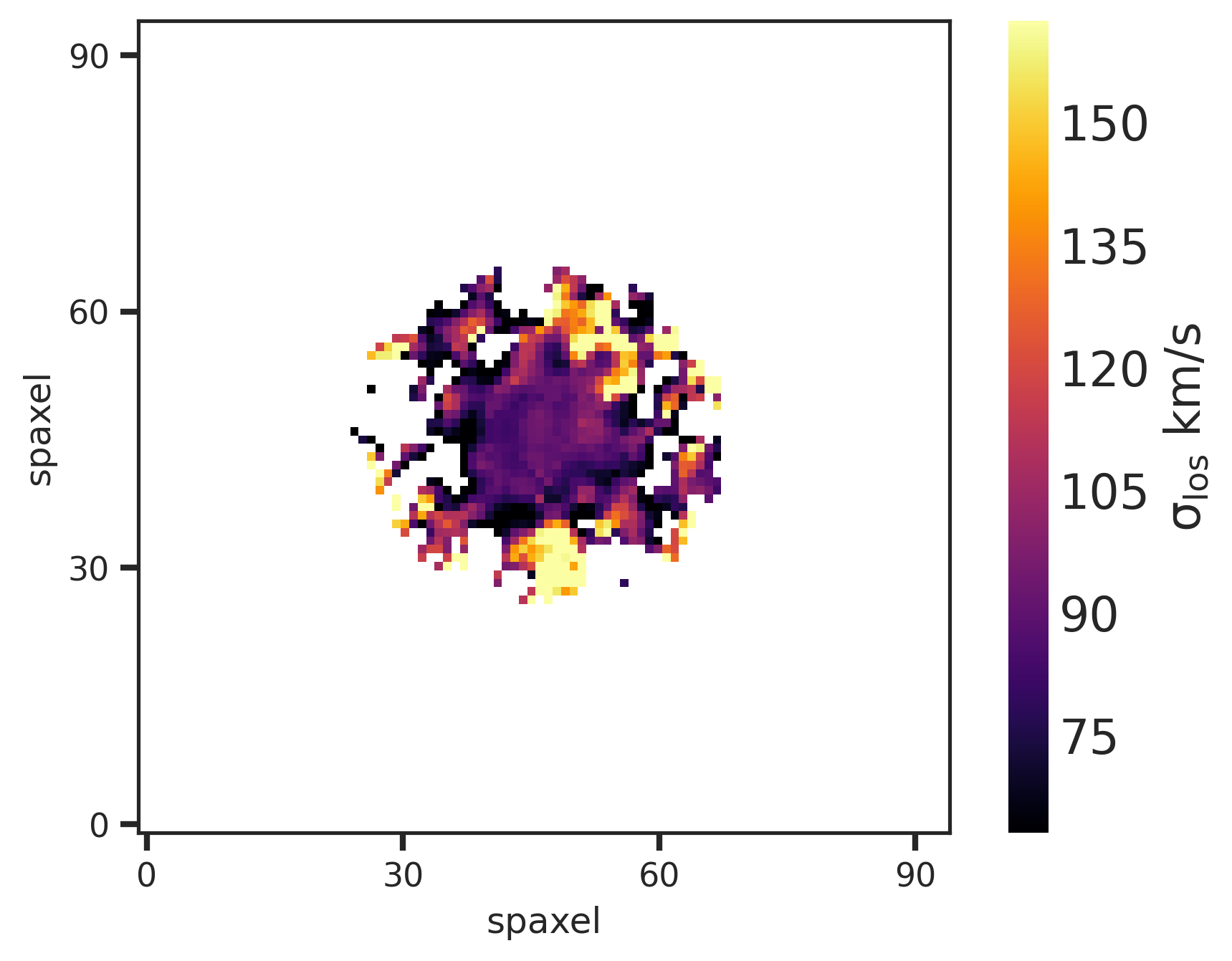}\vspace{0.02\textheight}
    
    \label{fig:ex_high_k1_low_k2}
    \caption{Two examples of galaxies (i.e. 9085-1902 top row, 11007-12705 bottom row) with high $k_1$ and low $k_2$. From left to right: SDSS \textit{gri} image, l.o.s. stellar velocity and l.o.s. stellar velocity dispersion. The magenta hexagon is the MaNGA field of view.}
    
\end{figure*}

9085-1902 is a galaxy with a clear rotational pattern where the IFU (i.e. the magenta hexagon in the top left panel of Fig.~\ref{fig:ex_high_k1_low_k2}) is probably not caching the whole galactic disc. In this case, the galaxy shows mild rotational support (i.e. $k_{\rm lum}\simeq 0.4$) modelled through a rotationally supported inner disc ($k_1\simeq0.9$) which primarily contributes to the galactic disc.

11007-12705 is a compact galaxy with $k_{\rm lum}\simeq 0.4$. In this case, the presence of a rotationally-supported inner disc can be understood by looking at the velocity field (i.e. second row, middle panel of Fig.~\ref{fig:ex_high_k1_low_k2}) which hints at some sort of regular rotation in the galaxy centre.  

\label{lastpage}
\end{document}